\documentclass[11pt]{article}
\usepackage{graphicx}
\usepackage{amssymb}
\usepackage{amscd}
\usepackage{mathrsfs}
\usepackage{longtable,lscape}
\usepackage{amsthm}
\usepackage{amsfonts}
\usepackage{amsmath}
\usepackage{bbm}
\usepackage{float}
\usepackage{url}

\newcommand{\real}{{\mathbb R}}
\newcommand{\captionfonts}{\footnotesize}
\makeatletter  
\long\def\@makecaption#1#2{%
  \vskip\abovecaptionskip
  \sbox\@tempboxa{{\captionfonts #1: #2}}%
  \ifdim \wd\@tempboxa >\hsize
    {\captionfonts #1: #2\par}
  \else
    \hbox to\hsize{\hfil\box\@tempboxa\hfil}%
  \fi
  \vskip\belowcaptionskip}
\makeatother 
\begin{document}
\title{On the Conceptuality Interpretation of Quantum \\ and Relativity Theories}
\author{Diederik Aerts$^1$, Massimiliano Sassoli de Bianchi$^{1,2}$, Sandro Sozzo$^{3}$  and Tomas Veloz$^{1,4}$ \vspace{0.5 cm} \\ 
        \normalsize\itshape
        $^1$ Center Leo Apostel for Interdisciplinary Studies, 
         Brussels Free University \\ 
        \normalsize\itshape
         Krijgskundestraat 33, 1160 Brussels, Belgium \\
        \normalsize
        E-Mails: \url{diraerts@vub.ac.be,msassoli@vub.ac.be}
	  \vspace{0.5 cm} \\ 
        \normalsize\itshape
        $^2$ Laboratorio di Autoricerca di Base, \\
         \normalsize\itshape
        Lugano, Switzerland
	  \vspace{0.5 cm} \\ 
        \normalsize\itshape
        $^3$ School of Management and IQSCS, University of Leicester \\ 
        \normalsize\itshape
         University Road, LE1 7RH Leicester, United Kingdom \\
        \normalsize
        E-Mail: \url{ss831@le.ac.uk} 
          \vspace{0.5 cm} \\ 
        \normalsize\itshape
        $^4$ Instituto de Filosof{\' i}a y Ciencias de la Complejidad IFICC, \\ 
         \normalsize\itshape
        Los Alerces 3024, \~Nu\~noa, Santiago, Chile
         \\
        \normalsize
        E-Mail: \url{tveloz@gmail.com}
       	\\
              }
\date{}
\maketitle
\begin{abstract}
\noindent
How can we explain the strange behavior of quantum and relativistic entities? Why do they behave in ways that defy our intuition about how physical entities should behave, considering our ordinary experience of the world around us? In this article, we address these questions by showing that the comportment of quantum and relativistic entities is not that strange after all, if we only consider what their nature might possibly be: not an objectual one, but a conceptual one. This not in the sense that quantum and relativistic entities would be human concepts, but in the sense that they would share with the latter a same conceptual nature, similarly to how electromagnetic and sound waves, although very different entities, can share a same undulatory nature. When this hypothesis is adopted, i.e., when a \emph{conceptuality interpretation} about the deep nature of physical entities is taken seriously,  many of the interpretational difficulties disappear and our physical world is back making sense, though our view of it becomes radically different from what our classical prejudice made us believe in the first place.
\end{abstract}

\section{Introduction\label{intro}}

In 1924, Luis de Broglie, in his PhD thesis \cite{De Broglie1924}, made one of the boldest moves in the history of modern physics. Following Planck and Einstein's introduction of a dual particle-like aspect associated with light waves, to ``explain'' their strange behavior in certain experiments, de Broglie, reasoning in a specular way, introduced the hypothesis that a wave-like aspect should also be associated with physical entities that, until that moment, were only considered to be corpuscles, like electrons, neutrons and protons. Like all new wild ideas, physicists were initially very unsure about the value of de Broglie's hypothesis, but fortunately Langevin had the foresight to send a copy of his thesis to Einstein, who was immediately conquered by the idea, so that de Broglie was ultimately granted his doctorate. The rest is history: a few years later, Davisson and Germer in the USA, and G.P. Thomson in Scotland, confirmed by means of diffraction experiments that electrons could also behave as waves. In 1929, Louis de Broglie was then awarded the Nobel Prize in physics for his discovery of the wave nature of electrons, which as we know laid the foundations of quantum mechanics, and in 1937 also Davisson and Thomson received the Nobel Prize, for their historical diffraction experiments. 

The aim of the present paper is to discuss about a more recent ``move \`a la de Broglie,'' which is also the result of a specular reasoning. The starting point is the new and booming research field known as \emph{quantum cognition}, where the mathematical formalism of quantum mechanics was applied with unexpected success to model human concepts and their interaction with human minds, showing that we humans think and take decisions pretty much in a quantum-like way. This doesn't necessarily mean that our brains would be like \emph{quantum computers}, exploiting the existence of quantum effects at the micro-level, but it certainly means that a quantum-like behavior is not the prerogative of the micro-entities, being instead a form of organization that can be found at different structural levels within our reality \cite{AertsEtal2015}. Now, if the human conceptual entities are to be associated with a quantum-like behavior, and therefore possess a quantum nature, one can introduce the hypothesis that, the other way around, the micro-physical (quantum) entities should also be associated with a conceptual-like behavior, and therefore possess a conceptual nature similar to that of the human concepts. However, different from the wave-particle duality, the quantumness-conceptuality binomial would not be the expression of a relation of complementarity, but rather of a relation of similarity, in the sense that \emph{quantumness} and \emph{conceptuality} would just be two terms pointing to a same reality, or nature, which can manifest at different organizational levels within reality.

The above hypothesis, that quantum entities are conceptual, was proposed by one of us in 2009 \cite{Aerts2009a,Aerts2010a,Aerts2010b,Aerts2013,Aerts2014}, and in the present work we will 
demonstrate its explicative power by reviewing some of the quantum  situations in which it has been applied so far, among those considered to be not yet fully understood, or even not understandable. We will do the same for the interpretational difficulties of special relativity theory, thus showing that the \emph{conceptuality interpretation} really represents a possible fundamental step forward in our understanding of the stuff our world is made of, and a candidate for the construction of a coherent framework for both quantum and relativity theories, and maybe also evolutionary theories \cite{AertsSassoli2017b}. But before doing so, some words of caution are necessary.  From the hypothesis that quantum entities would be conceptual entities carrying meaning and exchanging it with pieces of ordinary matter, a \emph{pancognitivist} view naturally emerges \cite{AertsSassoli2017b}, where everything within our reality participates in cognition, with human cognition being just an example of it, expressed at a specific organizational level. This, however, is not meant to be interpreted as an anthropomorphization of reality, because human cognition is to be considered as a much younger and hence still rather unsophisticated form of that more fundamental conceptual structure constituting the global reality. 

There is also no connection at all with any type of idealistic philosophical views, where physical theories would be considered to be mere theories of human mental content. Quite on the contrary, the conceptuality interpretation is a genuine realistic view, where conceptual entities are seen as entities that can be in different states and be subjected to measurement processes, which are processes not only of \emph{discovery} (of the properties that were already actual), but also of \emph{creation} (of those properties that were only potential prior to the measurement and could become actual through its execution).  Also, the conceptual substance forming our global reality is not even necessarily connected to the human cognition, its existence being certainly independent of it, i.e., even when us humans, as a cognitive species, had not yet come into existence on the surface of planet Earth, the fundamental conceptual substance forming our global reality was already there, because also its quantum aspects were already there. To make even more clear the realistic stance underlying our conceptuality interpretation, if the dinosaurs would not have become extinct (probably due to the impact of an asteroid) and would have further evolved their cognitive talents, they could easily have been them the first to explore the conceptual layer existing within their species, in a similar way as they might have also have been the first to discover the quantum nature of the micro world.

Having said this, and before proceeding in the next sections by describing how the conceptuality interpretation can explain different quantum and relativistic phenomena, it is interesting to reflect for a moment about the reasons why quantum physics has remained so far so difficult to understand, which is also the reason why so many interpretations have seen the day since it was fully formulated in the thirties of the past century. The case of relativity theory is only apparently different, as the majority of physicists seem to cultivate the belief that relativity would be well understood, or at least much better understood than quantum mechanics, which in our view is only the fruit of a misconception, as we will emphasize later in the article. A first important point to consider is that the very fact that numerous quantum interpretations still exist today can be seen as the sign that none of them has been able to provide so far those notions that would capture, in its entirety, the reality that quantum theory aims to describe, and therefore obtain a general consensus. We believe that one of the reasons for their failure is the fact that most of them only try, somehow nostalgically, to interpret the mathematical quantum formalism in terms of classical spatiotemporal notions.

To better explain what the fulcrum of the problem is, when one tries to understand quantum (and relativistic) entities, let us use a metaphor. During the eighteenth century, the first British settlers who landed on the Australian continent were confronted with a totally new territory, both for the uses and customs of the natives, the Aborigines, and for the mysterious flora and fauna that populated those distant lands. Among Australian animals there was one in particular that struck the imagination of the settlers. Every now and then they could see it in the vicinity of the watercourses, but being shy it was difficult to see it clearly. When they could have a glimpse of it from the front, seeing its flat beak and its two palmed feet, they probably exclaimed: ``It's a duck!'' But then, when it turned around and ran away, they realized that it had not two, but four paws, and a dense fur. So, they probably also exclaimed: ``No, it's a mole!'' And by dint of exclaiming that: ``It's a duck!... No, it's a mole!... No, it's a duck!... No, it's a mole!..." in the end they decided to call it a \emph{duckmole}! (Our little story is of course a caricature). In other words, they baptized this odd animal with a paradoxical name, obtained by the composition of the names of two different animals. Such a designation, of a dualistic nature, was clearly only provisional, since no animal can simultaneously be a duck and a mole, and when they finally managed to observe it more closely and more attentively, they realized that it was neither, but something completely different, so finally the animal got a name of its own: \emph{platypus}!\footnote{Prior to the arrival of the European settlers, Aboriginal people had many names for the animal, including \emph{boondaburra}, \emph{mallingong} and \emph{tambreet}. The first scientific description of the platypus (\emph{ornithorhynchus anatinus}) is attributed to the English botanist and zoologist George Shaw, whose first reaction was to believe the specimen to be a hoax, made of several animals sewn together.}

The above curious anecdote was used by Jean-Marc Levy-Leblond \cite{Levy-Leblond1999} to illustrate the situation of physicists at the beginning of the past century, who like the Europeans settlers were confronted with entities -- the microscopic ones, such as photons and electrons -- whose appearance could change depending on the experimental settings, sometimes being observed as particles (moles) and other times as waves (ducks). And again by dint of exclaiming that: ``It's a particle!... No, it's a wave!... No, it's a particle!... No, it's a wave!..." in the end they also decided to provisionally denote them \emph{waveparticles}, \emph{wavicles}, etc, \cite{Bunge1973,LeblondBalibar1984}, i.e., to talk about them in terms of a \emph{wave-particle duality}. But in the same way a platypus is neither a duck nor a mole, and certainly not simultaneously a duck and a mole, a microscopic quantum entity is also neither a particle nor a wave, and certainly not simultaneously a particle and a wave. The waveparticle dualistic designation is in fact only the result of a fleeting observation of their behavior, and if one takes the time to observe them with more attention, it becomes clear that what they truly are is ``something else,'' something completely different from the discrete and local notion of a particle as well as from the continuous and extended notion of a wave, since both of these notions are spatial, while one of the most salient features of the microscopic quantum entities is precisely that of not being representable as entities permanently present in space (or spacetime). In other words, we know what quantum entities certainly are not: they are \emph{non-spatial} entities (and more generally, as we are going to also discuss, \emph{non-spatiotemporal} entities). 

However, knowing what a microscopic quantum entity is not, does not tell us what it is, i.e., what its nature truly is. The same was true for the previous example of the platypus: knowing what it was not, was not sufficient to determine its nature, which is the reason why a controversy lasted for quite some time among European naturalists, when they discovered the unusual characteristics of the animal.\footnote{Today the platypus is classified as a \emph{monotreme}: a mammal that can lay eggs, with the male also having a spur on the hind foot that delivers a venom capable of causing severe pain to humans, and with many other structural differences compared to common mammalians.} Understanding the nature of a quantum entity is fundamental because the behavior of a physical entity can appear to us very strange, if not incomprehensible, if we believe it is something that it is not, whereas its behavior may all of a sudden become perfectly normal and fully understandable if we can correctly identify its nature. In that respect, it is important to emphasize that a physical theory requires not only a mathematical formalism, but also a network of physical concepts coherently relating to the latter and capable of providing a meaningful physical representation of the reality the theory aims to describe \cite{De Ronde2016}. And of course, among these physical concepts the most crucial one is that identifying the nature of the physical entities the theory is about. For instance, before the advent of quantum mechanics, the concept of \emph{particle} (or corpuscle) was fundamental in order to make sense of the other notions associated with the theory (of classical mechanics), like those of position, velocity, mass, etc., which in turn were associated with specific mathematical objects in the formalism. 

So, to make sense of quantum mechanics, the first thing one needs to do is to find a  notion  specifying what the nature of a micro-physical entity is. We know it is not a particle notion, or a wave notion, nor a waveparticle notion,  so, what is it? The standard answer is that we don't have nothing valid at hand 
to represent the nature of a quantum entity, but, that's it? As Arthur Conan Doyle used to point out more than once, in his Sherlock Holmes stories, sometimes the best place to hide something is to keep it in plain sight. And according to the conceptuality interpretation, what has always been in plain sight, but precisely for that was very hard to notice, is that the notion one should use to represent the nature of a quantum entity, and make full sense of its behavior, is the very notion of \emph{concept}! In other words, human concepts would not be the only category of conceptual entities with which we humans have interacted: the so-called microscopic quantum entities would form another category of conceptual entities, much more ancient and structured than our human ones, and as soon as we reset our mental parameters and start thinking of, say, an electron, not as an object but as a conceptual entity,  most of  the mystery of its quantum behavior disappears, as we are now going to show by considering different physical situations.

\section{The double-slit experiment\label{double}}

\begin{figure}[htbp]
\begin{center}
\includegraphics[width=14cm]{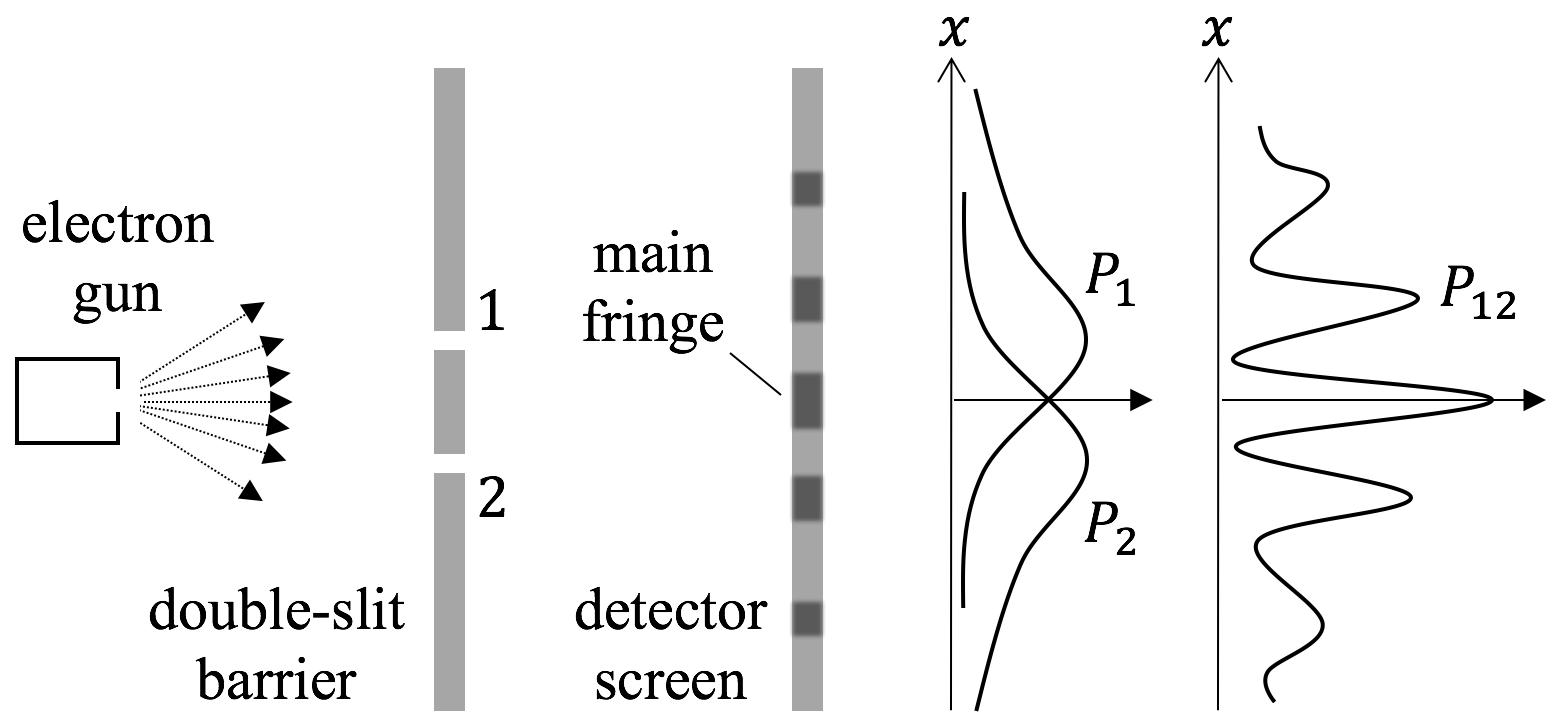}
\caption{In the double-slit experiment an electron source fires the electrons towards a barrier having two slits. If slits 1 is open and slit 2 is closed, the probability distribution for detecting an electron at a distance $x$ from the center of the detection screen is $P_1$. If slits 2 is open and slit 2 is closed, the probability distribution is $P_2$. If both slits are open, the probability distribution $P_{12}$ is not proportional to the sum of $P_1$ and $P_2$, as one would expect if the electrons were particles, but is a more complex function describing a fringe interference pattern, with the main fringe being the one at the center of the detection screen.}
\label{Figure1}
\end{center}
\end{figure}

Richard Feynman used to say that the double-slit experiment has in it the heart of quantum mechanics and contains the only mystery. Certainly, it contains part of the mystery, so let us start by describing this experiment to show how it can go away, if we only start thinking  of the micro-physical entities interacting with double-slit barrier -- let us assume they are electrons -- not as particles, or waves, but as conceptual entities. For this, we begin by recalling why the double-slit experiment is impossible to explain in any classical way. The reason is simple: the localized impacts on the detector screen seem to show that the entities in question are particle-like. On the other hand, the fringe pattern one observes, when multiple impacts are collected, reveals  that what traverses the double-slit is more like a wave phenomenon, able to create interference effects (see Figure~\ref{Figure1}). And since a wave is not a particle, and \emph{vice versa}, the observed behavior of the electrons cannot be consistently explained.

More precisely, if they would be like small projectiles, then a \emph{compositional} interpretation of the experiment should be possible, with the pattern of impacts obtained when both slits are open being deducible from the patterns of impacts obtained when these are opened one at a time, instead of simultaneously. This means that the probability $P_{12}(x)$ of having an impact at a point $x$ of the detection screen, in the situation where the two slits are open, should be given by the uniform average of the probabilities $P_1(x)$ and $P_2(x)$ of having an impact at that same point when only slit 1 or only slit 2 are open, respectively, i.e., $\bar{P}_{12}(x)={1\over 2}[P_1(x) + P_2(x)]$. But since we generally have ${P}_{12}(x)\neq\bar{P}_{12}(x)$, even though the electrons \emph{appear} to be corpuscular, as they leave traces on the screen in the form of point-like impacts, they cannot be such, as the obtained complex fringe pattern demonstrates. Note that the one-slit probability distributions $P_1(x)$ and $P_2(x)$ are compatible with the hypothesis that the electrons would be entities of a corpuscular nature. It is really when both slits are simultaneously open that the distribution of impacts on the detection screen becomes incompatible with the corpuscular assumption, being no longer deducible as a uniform average of the one-slit distribution probabilities. Reasoning in probabilistic terms, there will be points $x$ on the detection screen where the probability of observing an electron will differ sensibly from the value given by the uniform average $\bar{P}_{12}(x)$, in the sense that there will be points of \emph{overexposure} [corresponding to a probability \emph{overextension}: ${P}_{12}(x)>\bar{P}_{12}(x)$], and points of \emph{underexposure} [corresponding to a probability \emph{underextension}: ${P}_{12}(x)<\bar{P}_{12}(x)$], meaning that one has to correct the uniform average by introducing a third term $I(x)$, an \emph{interference contribution} responsible for these overextension (constructive interference) and underextension (destructive interference) effects: $P_{12}(x)=\bar{P}_{12}+I(x)$.

Let us now consider the hypothesis that the electrons are conceptual entities, i.e., entities behaving in a way which is similar to how human concepts behave. And let us also assume that the measuring apparatus, and more specifically the screen detector, is an entity sensitive to the meaning carried by the electrons and able to answer questions when the latter are addressed in operational terms, i.e., by enacting them through the construction of a specific experimental arrangement. Of course, the screen detector mind-like entity does not speak our human language, and will only communicate by means of signs that are the electrons' traces of impact on its surface, which we have to correctly interpret, and for this we have to understand what is the meaning that is attached to the impacts appearing in the different positions. Now, the questions the screen-mind entity is possibly answering, by means of its ``pointillistic language,'' are here the following three: (a) ``What is a good example of an impact point of an electron passing through slit 1?'' (b) ``What is a good example of an impact point of an electron passing through slit 2?'' (c) ``What is a good example of an impact point of an electron passing through slit 1 or 2?'' These three questions can be addressed in practical terms by having only slit 1 open, only slit 2 open, and both slits open, respectively. Of course, the electron conceptual entity will then be in a state that depends on the configuration of the barrier. When only slit 1 is open, it will be in a state $\psi_1$, corresponding to the conceptual combination \emph{The electron passes through slit 1}. When only slit 2 is open, it will be in a state $\psi_2$, corresponding to the conceptual combination \emph{The electron passes through slit 2}. And when both slits are open, it will be in a state $\psi_{1,2}$, corresponding to the conceptual combination \emph{The electron passes through slit 1 or 2}.\footnote{The notion of ``passing through'' remains a very human way of conceptualizing the question addressed to the measuring apparatus. Indeed, when we say ``passing through,'' or even ``impact point,'' we are already attributing to the electron spatial properties that it does not necessarily have. In other words, we are already looking at things from the bias of our spatial prejudices. On the other hand, if ``passing through'' is more generally understood as a way to express the fact that the only regions of space occupied by the barrier where there is a zero probability of absorbing the electrons are those of the two slits (when they are open), then the notion of ``passing through'' can still be used to conveniently describe the experiment in a way that our human minds can easily understand. A more general and probably more correct way of formulating the above three questions would be: (a) ``What is a good example of an effect produced by an electron interacting with the barrier having only slit 1 open?'' (b) ``What is a good example of an effect produced by an electron interacting with the barrier having only slit 2 open?'' (c) ``What is a good example of an effect produced by an electron interacting with the barrier having both slit 1 and 2 open?''}

If the above states are represented by complex vectors in a Hilbert space, one can easily recover the interference pattern at the detection screen by representing $\psi_{1,2}$ as a normalized superposition of $\psi_1$ and $\psi_2$, i.e., $\psi_{1,2}={1\over\sqrt{2}}(\psi_1+\psi_2)$. Then, the probability density $P_1(x)$ [resp., $P_2(x)$] that the screen-mind provides the answer $x$ to question (a) [resp., (b)] is $P_1(x) = |\psi_1(x)|^2$ [resp., $P_2(x) = |\psi_2(x)|^2$], whereas the probability density for $x$ to be selected as a good example of an electron passing through slit 1 or 2 [question (c)] is: $P_{12}(x)=|\psi_{1,2}(x)|^2 = {1\over 2}|\psi_1(x)+\psi_2(x)|^2 = {1\over 2}[|\psi_1(x)|^2+|\psi_2(x)|^2] +2\Re\, \psi_1^*(x)\psi_2(x)$, where $I(x)=2\Re\, \psi_1^*(x)\psi_2(x)$ is the interference contribution, accounting for the overextension and underextension effects, and the symbol $\Re$ denotes the real part of a complex number. This is of course the well-known quantum mechanical rule saying that when we are in the presence of alternatives (slit 1 or 2), the probability amplitude is obtained by the normalized sum of the probability amplitudes for the alternatives considered separately. But what we want now to understand is the emergence of this fringe pattern from the conceptuality hypothesis viewpoint. In other words, we want to understand the cognitive process operated by the detection screen, when viewed as a mind-like entity answering the above three questions. 

First of all, we have to observe that such cognitive process cannot be deterministic. Indeed, the specification ``passing through a slit" is not sufficient to describe a unique trajectory in space. This is so also because being the electron a conceptual entity, it cannot be attached with \emph{a priori} spatial properties. These will have to be acquired by interacting with the apparatus, so as to give a sense to the very notion of ``passing through." And since there are many ways in which a spatial entity is able to pass through a slit, the screen-mind will have to choose from among several possibilities, and choosing one among these possibilities is a \emph{symmetry breaking} process whose outcomes cannot be predicted in advance, which is the reason why every time the question is asked the answer (the trace of the impact on the screen) can be different. However, answers cannot be totally arbitrary, as is clear that the question specifies that the electron passes, for example in case of question (a), through slit 1. So, the screen-mind will certainly manifest a greater propensity to respond by means of an impact point located in a position in proximity of slit 1, which means that the symmetry breaking process will be a weighted one, with some outcomes having greater probability than others (more will be said about measurements in Sec.~\ref{measurement}). Of course, things get more interesting when we consider question (c), as in this situation not only there are many possibilities about how the electron will pass through either slit, but also about which slit, 1 or 2, it will pass through. Confronted with this situation, the screen-mind will thus have to select those answers that best express this double level of uncertainty, producing an impact point that will be typical of an electron conceptual entity having acquired spatial properties and passing through slit 1 or 2. And when the question is operationally asked several times, the result will be the typical fringe structure shown in Fig~\ref{Figure1}.

Let us delve into the screen-mind to try to understand how such fringe structure can emerge. For this, let us concentrate on its most salient feature: the central fringe, which is the one with a higher density of impacts, located at equal distance from the two slits. This is where the screen-cognitive entity is most likely to manifest an answer, when subjected to question (c). To understand why, we can observe that an impact in the region of the central fringe corresponds to a situation of maximum doubt regarding the slit the electron would have used to cross the barrier, or even the fact that it would necessarily have passed through one or the other slits, in an exclusive manner. Therefore, it constitutes a perfect exemplification, in the form of an impact point on the screen, of the concept ``an electron passing through slit 1 or 2." Now, if the region in between the two slits is a region of overextension, the two regions opposite the two slits are instead regions of underextensions, showing a very low density of impacts. To understand why, we can observe that an impact in the regions facing the two slits would not make us doubt about the slit through which the electron has passed. In other words, an impact point in the two regions opposite the slits would constitute a very bad exemplification of the concept ``an electron passing through slit 1 or 2." Moving then from these two regions away from the center, we will be back again in a situation of doubt, although less perfect than that expressed by the central region, so regions of overextension will manifest again, but this time less intense, and then again regions of underextension will come, and so on, producing in this way the typical fringe pattern observed in experiments.

Considering the above conceptuality explanation of the double-slit experiment, we see that the wave aspect associated with electrons (mathematically described by the wave function $\psi_{1,2}$, evolving according to the Schr\"{o}dinger equation), is just a convenient way to model, by means of constructive and destructive interference effects, the different overextension and underextension effects that result from the cognitive (symmetry breaking process) through which a good (concrete) exemplar for an abstract conceptual entity is each time provided, when the interrogative context forces the electronic conceptual entity to enter the spatiotemporal theater, by means of a localized impact on the screen. Of course, this impact should not be mistaken as a trace left by a corpuscular entity with a well-defined trajectory in space, as it will be better explained in the following sections. Now, to confer more credibility to the above narrative, and considering that an electron and a human concept are assumed to share the same conceptual nature (in the same way an electromagnetic wave and an acoustic wave, even though they are different physical phenomena, can share the same wavy nature), one should be able to also show that human minds are able of producing similar interference figures, when subjected to interrogative contexts that confront them with genuine alternatives. This is indeed the case: human minds, when interacting with concepts, will generally produce overextension and underextension effects having a very complex pattern, in fact much more complex (less symmetric) than those produced by screen-minds interacting with electrons (or photons). Let us very briefly describe an experiment where this has been explicitly demonstrated, referring the interested reader to \cite{Aerts2009a,AertsSassoli2017} for the details.

\begin{figure}
\begin{center}
\includegraphics[width=14cm]{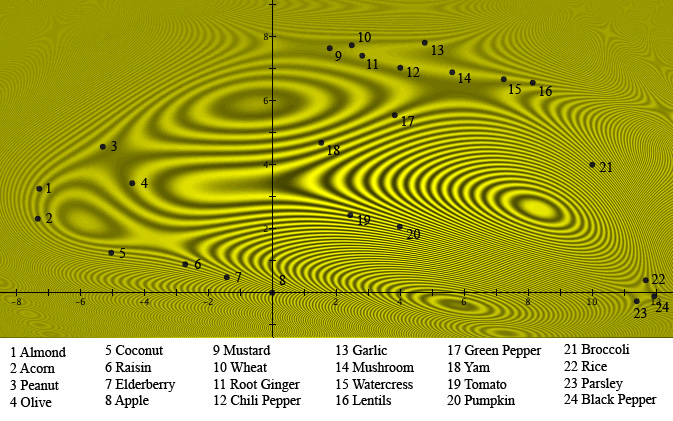}
\end{center}
\vspace{-15pt}
\caption{The interference-like figure describing the overextension and underextension effects contained in Hampton's data, when the participants had to select exemplars representative of the disjunction \emph{Fruit or vegetable}. For more details about how this figure was obtained, see \cite{Aerts2009a}.}
\label{Figure2}
\end{figure}

In the eighties of last century, the cognitive psychologist James Hampton conducted an experiment where 24 exemplars of \emph{Food}\footnote{These are: \emph{Almond, Acorn, Peanut, Olive, Coconut, Raisin, Elderberry, Apple, Mustard, Wheat, Ginger root, Chili pepper, Garlic, Mushroom, Watercress, Lentils, Green pepper, Yam, Tomato, Pumpkin, Broccoli, Rice, Parsley, Black pepper.}} were submitted to 40 students, asking them if they were typical (i.e., good examples) of a (a) \emph{Fruit}; (b) \emph{Vegetable} or (c) \emph{Fruit or vegetable} \cite{Hampton1988}. These different exemplars of \emph{Food} play here the same role as the different locations $x$ on the detection screen, in the double-slit experiment, with the concept \emph{Fruit} (resp. \emph{Vegetable}) playing the role of slit 1 (resp., slit 2). If the decision-making process of the students, when subjected to question (c), would be of a sequential kind (they first choose between \emph{Fruit} and \emph{Vegetable} and then, if they chose the former, they select a good example of \emph{Fruit}, and if they chose the latter, they select a good example of \emph{Vegetable}), then the probability of selecting a given exemplar of \emph{Food} should correspond to the uniform average of the probabilities describing the situations of questions (a) and (b). But this is not what Hampton's data reveal, which contain instead a complex pattern of overextension and underextension effects. When these data are represented in a quantum-like way, using two two-dimensional functions interpolating the outcomes of questions (a) and (b), then a normalized superposition of these two functions to interpolate the data of question (c), a complex interference figure is revealed, reminiscent of those obtained in the phenomena of birefringence \cite{Aerts2009a,AertsSassoli2017} (see Figure~\ref{Figure2}). 

We conclude this section by an important remark. In our discussion, we made a distinction between the detector screen, playing the role of the structure sensitive to the meaning carried by the electrons, and the barrier, playing the role of the structure allowing the three questions (a), (b) and (c) to be addressed in operational terms, when slit 1, slit, 2, and both slits are open, respectively. This distinction is however not fundamental and was just used to obtain a stronger analogy with our typical human experience, when we distinguish the mind answering a question from the process of addressing a question to it, for instance orally or in writing. In fact, the entire structure of the experimental apparatus should really be interpreted as the mind-like entity, as is clear that not only the screen but also the other material parts, in particular the barrier, interact as a whole with the electrons' conceptual entities. So, a more correct image consists in saying that the structure of the entire apparatus mind-like entity changes depending on the question that is being asked. More precisely, the effect of asking question (a) [resp., (b) and (c)] is the opening of slit 1 (resp., slit 2 and both slits) at the level of the barrier, and the actual answering of the question is the process of having the electron conceptual entity entering it and leaving a trace on the detector screen. 

Having analyzed the double-slit experiment, we want to consider in the next section another paradigmatic quantum experiment that remains impossible to understand if one does not give up the prejudice that the micro-physical entities would be particles or waves, i.e., spatiotemporal phenomena, and  becomes instead very easy to explain if one assumes that they are conceptual (meaning) entities.

\section{Delayed-choice experiment\label{delayed}}

\begin{figure}
\begin{center}
\includegraphics[width=11cm]{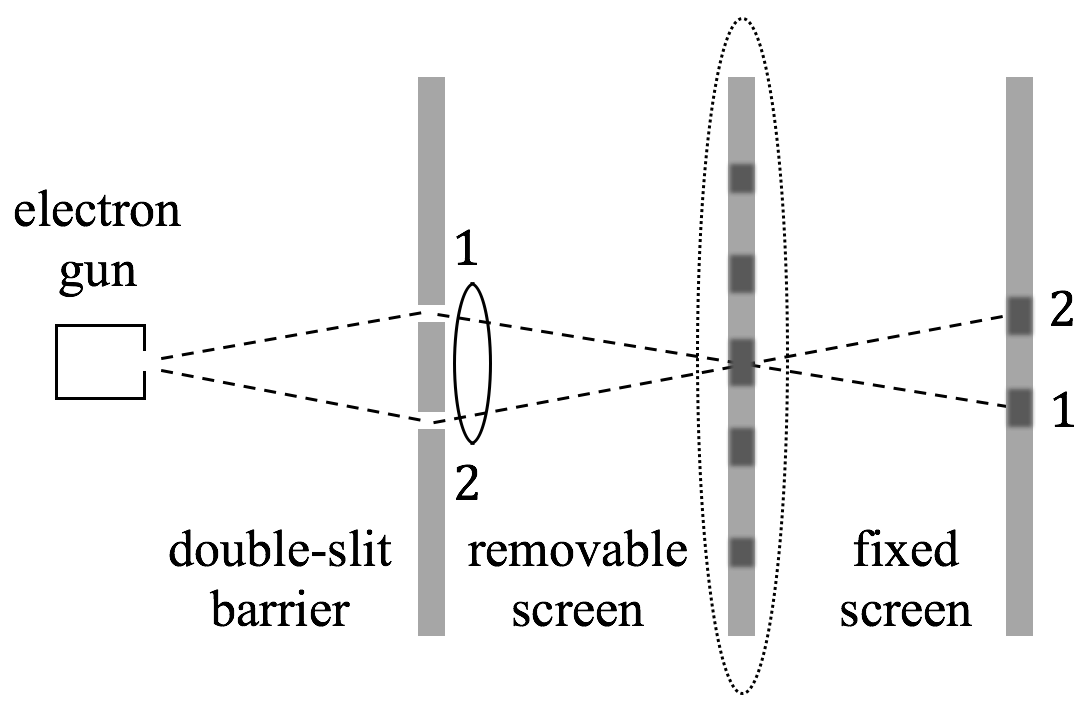}
\end{center}
\vspace{-15pt}
\caption{A schematic diagram of a delayed choice experiment, where one of the two detection screens is removable, so that a wave-like or particle-like context can be created by either leaving it in place or removing it, respectively. The lens element close to the double-slit barrier makes the wave function components coming from the two slits to slightly diverge, so there will be relevant interference effects only at the location of the removable screen, but not at the more distant location of the fixed screen. This means that the latter will not show a fringe pattern, but only two distinct and equivalent regions of impact, which can be associated with electrons emerging either from slit 1 or from slit 2.}
\label{Figure3}
\end{figure}

In 1978, Wheeler considered the following experiment \cite{Wheeler1978}. A quantum entity, say an electron, enters an apparatus like the previously described double-slit one, with the difference that its arrangement can be changed at the last moment, before the electron is finally detected. The variable arrangements that are considered are two: a wave arrangement, like the one used in a typical double-slit experiment, which gives rise to overextension and underextension effects, and a particle arrangement, corresponding to the situation where the detection screen is removed and replaced by a second detection screen, located at a greater distance, so that the impacts detectable on it become compatible with a classical particle-like description (no overextension or underextension effects); see Figure~\ref{Figure3}.
More precisely, since the apparatus cause the wave function's components coming from the two slits to diverge, they will not anymore superimpose when they arrive at the place where the second (fixed) screen is present, so that the traces of the impacts on it allow to determine with no ambiguity the wave function component they are associated with, i.e., which path was followed by the electron, if interpreted as a particle. The experimental setting is however such that the arrangement can be changed extremely rapidly, and the result of the many experiments so far conducted is that though the arrangement is changed at the very last moment, the electrons (or any other micro-physical quantum entities) behave as if it was present since the very beginning. 

Experiments of this kind (see for instance \cite{Jacquesetal2007}) demonstrate the inadequacy of the wave-particle duality. As a matter of fact, if the electron quantum entity would behave as a wave (i.e., as a spatial entity passing through both slits) or as a particle (i.e., as a spatial entity passing either through slit 1 or slit 2), depending on the experimental arrangement, then, when the latter is changed at the last moment, the electron (assumed to be an entity propagating through space) should have left already the double-slit barrier region, and the delayed change should not be able to affect its prior wave or particle behavior. This however is not what is observed in the experiments, where everything happens as if the electron would have ``delayed its choice'' (from which the name that was given to these experiments) of manifesting either as a wave-like phenomenon or as a particle-like phenomenon, until the final arrangement is decided. Facing the implications of these experiments, Wheeler famously affirmed the following \cite{Wheeler1978}: ``Then let the general lesson of this apparent time inversion be drawn: ‘No phenomenon is a phenomenon until it is an observed phenomenon.’ In other words, it is not a paradox that we choose what shall have happened after ‘it has already happened.’ It has not really happened, it is not a phenomenon, until it is an observed phenomenon.''

If by ``phenomenon'' we understand a ``spatial phenomenon,'' then we can only agree with Wheeler's statement, which indicates that we cannot understand the behavior of an electron by depicting it as a spatial entity, be it a wave, a particle, or a waveparticle. In other words, what these experiments show is that electrons, and any other micro-physical entities, are \emph{non-spatial} entities: when the gun fires an electron towards the double slit barrier, one should not imagine it as a wave or a particle propagating in space, but as a more abstract entity that is only drawn into space at the moment of its actual detection, either by the removable screen or by the fixed screen, depending on the final selection. Of course, the electron exists also prior to its detection, though not as an entity having already acquired spatiotemporal properties. Again, this is typical of the behavior of a conceptual entity whose state can change from a more abstract to a more concrete one, when interacting with a (mind-like) structure sensitive to the meaning it carries. 

Let us consider once again the conceptuality hypothesis, to see how the apparent delayed choice behavior of the electron becomes not only perfectly understandable, but also corresponds to what we would expect. As described in the previous section, the question that is being asked is: ``What is a good example of an impact point of an electron passing through slit 1 or 2?'' An answer to this question will be manifested either by the removable screen-mind, if maintained in place, or by the fixed screen-mind, if the former has been removed. These two cognitive entities, however, will encounter the electrons' conceptual entities in different states, because of their distinct spatial locations. From the perspective of the removable screen, which is closer to the double-slit barrier, the converging lens has no relevant effects, so the state $\psi_{1,2}$ of the electrons can be conveniently described by the conceptual combination: \emph{The electron passes through slit 1 or 2}. On the other hand, since the converging lens produces a relevant effect for the farther away fixed screen, it will interact with the electrons in a different state $\psi'_{1,2}$, which can be described by the conceptual combination: \emph{The electron passes through slit 1 or 2 and is subsequently strongly deviated from its trajectory by a converging lens}. These states being different, the meaning carried by the electron in the two situations is also different, so that the removable screen-mind will answer the question in the way described in the previous section, with a complex fringe pattern having a central major fringe, whereas the fixed screen will answer by randomly considering either an upper spot, associated with slit 2, or a lower spot, associated with slit 1 (see Figure~\ref{Figure3}). 

But why now a central spot is not anymore a good exemplar for expressing the doubt regarding which slit an electron has passed through? The reason is simple to understand: because of the presence of the converging lens, and the distance of the fixed screen, the state $\psi'_{1,2}$ of the electron can now be described, in more synthetic terms, by the conceptual combination: \emph{The electron passes through slit 1 either/or 2}. In other words, the ``or'' has been replaced by an exclusive-or (xor), conveying the meaning that the electron can pass through slit 1 or slit 2, but not through both of them. So, the fixed screen has to answer the same question of the removable screen, but with the additional information that the electrons do not pass through both slits simultaneously. This means that a central point on the screen will not be anymore a good example of the situation, as a central point expresses a much deeper form of doubt: one where not only we don't know the slit through which the electron has passed through, but also if it has passed through only one of them or both of them. Now, since the slit through which the electron passes through remains unspecified, the only option for the fixed screen-mind, to answer consistently, is to produce a point impact either in a location compatible with the situation of an electron passing through slit 1, 50\% of the times, or in a location compatible with the situation of the electron passing through slit 2, the other 50\% of the times, which is exactly what is observed in experiments. Using again the Hilbert space formalism, we now have: $P'_{12}(x)=|\psi'_{1,2}(x)|^2 = {1\over 2}|\psi'_1(x)|^2+{1\over 2}|\psi'_2(x)|^2$, i.e., the two alternatives are non-interfering, compatibly with a classical (compositional) description.

\section{Heisenberg's uncertainty relations\label{Heisenberg}}

Coming back for a moment to the wave-particle duality, and assuming that an interference pattern would be indicative of a wave, and the absence of it would be indicative of a particle, experiments like the one described in the previous section are usually interpreted by saying that the behavior of a quantum entity, like an electron, is determined by the type of measurement we perform on it. This is certainly correct, but only if we understand that the determination arises in the moment the quantum entity is actually detected, and not before, and this also means that if we do not want to abandon a realistic view of our physical reality, we have to accept that a micro-physical entity, prior to its detection, is usually neither in a wave nor in a particle state, but in a condition that cannot be associated with any specific spatial property. The \emph{de Broglie-Bohm theory} can certainly offer an alternative description here, as it assumes that a quantum entity is the simultaneous combination of both aspects: a particle and a (pilot) wave~\cite{Norsen2006}. However, if considered as a tentative to preserve spatiality, the theory, as is well-known, faces a serious problem when dealing with more than a single entity, as the pilot wave (or quantum potential) cannot then be described as a spatial phenomenon, hence the interpretational problem remains, and in a sense get even worse. 

If we understand conceptual entities as meaning entities that can be in different states (each state specifying the actual meaning carried by the conceptual entity), which can change either in a predictable way, when they are subjected to deterministic contexts, or in an unpredictable way, when they are subjected to indeterministic ones, like interrogative (measurement) contexts, it immediately follows that, by definition of what a state is, a conceptual entity in a given state cannot be at the same time in another, different state. We are of course stating the obvious, but this is really what is at the foundation of Heisenberg's uncertainty principle. Consider the human concept \emph{Animal}. When we use a single word to indicate this concept, we can say that it describes the most abstract of all its states, associated with a perfectly neutral (tautological) context, just conveying the meaning that: \emph{The animal is an animal}. Let us look right away at a parallel between the human concept \emph{Animal} and a micro-physical entity like an electron, which according to the conceptuality interpretation also possesses a conceptual nature. Non-relativistic quantum theory does not describe in formal terms the state of an electron in the condition of just being an electron. We usually describe an electron in contexts were the electron has already acquired some more specific properties, which in the theory are mathematically described by the given of a (Hilbert space) vector, or a density matrix. 

\begin{figure}
\begin{center}
\includegraphics[width=14cm]{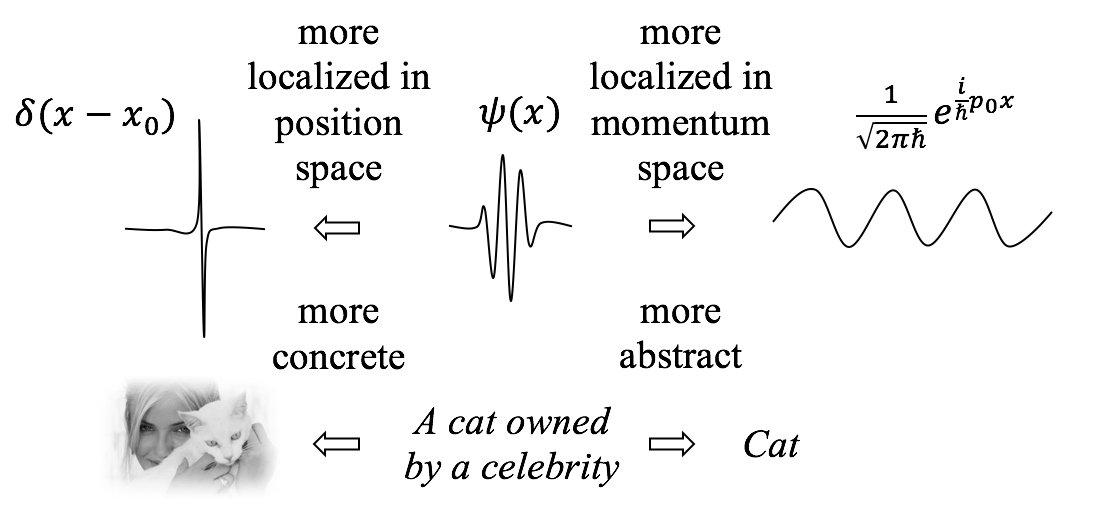}
\end{center}
\vspace{-15pt}
\caption{A schematic diagram describing the localized in space versus localized in momentum (opposite) directions which, according to the conceptuality interpretation, correspond to the concrete versus abstract directions.}
\label{Figure4}
\end{figure}

Consider then the following concepts: \emph{Dog}, \emph{Cat}, \emph{Horse}, etc. They are all specific examples of \emph{Animal}, hence, they specify different possible states of the animal-concept, and more precisely the states conveying the meanings: \emph{The animal is a dog}, \emph{The animal is a cat}, \emph{The animal is a horse}, etc. In other words, the concept \emph{Animal} can be in different states and the above are of course still examples of very abstract states, if compared to states that are determined by contexts that put for instance the \emph{Animal} concept in a one-to-one relation with a well-defined entity of our spatiotemporal theater. So, the conceptual combinations: \emph{The Labrador dog named Esmerelda owned by actress Anne Hathaway}, \emph{Cameron Diaz' white cat named Little Man}, \emph{The race horse named Lexington who set a record at the Metaire Course in New Orleans}, etc., are much more concrete states of the concept \emph{Animal}. A concept can thus be in different states, but certainly cannot simultaneously be in two different states, and some states are maximally abstract, others maximally concrete, and in between there are states (the majority of them) whose degree of abstraction is intermediary, like for instance the state described by the conceptual combination: \emph{A cat owned by a celebrity} \cite{Mervisetal1981,Rosch1999}. This means that a concept cannot be in a state that is maximally abstract and at the same time maximally concrete, and this is nothing but the conceptuality version of Heisenberg's uncertainty principle. (see Figure~\ref{Figure4}).

In the case of an entity like an electron, a maximally concrete state corresponds to the electronic entity being maximally localized in our three-dimensional space, while a maximally abstract state corresponds to it being maximally delocalized, i.e., to an electron being maximally localized in momentum space. In handbooks of quantum mechanics, Heisenberg uncertainty principle is usually stated by using the standard deviations of two non-commuting observables, like the position $q$ and momentum $p$ observables. The typical result is that the product $\sigma_q \sigma_p$ of their standard deviations must be bounded from below by a given value, for instance $\hbar\over 2$. The standard deviation $\sigma_q$ has here to be interpreted as a measure of the degree of concreteness of the state in which the electron micro-entity is, with $\sigma_q=0$ corresponding to a condition of maximum concreteness (i.e., maximum localization in position space) and $\sigma_q=\infty$ of minimum concreteness. Similarly, $\sigma_p$ has to be interpreted as a measure of the degree of abstractness of the state in which the electron is, with $\sigma_p=0$ corresponding to a condition of maximum abstractness (i.e., maximum localization in momentum space) and $\sigma_p=\infty$ of minimum abstractness. It is then clear that the product $\sigma_q \sigma_p$ must be bounded from below, as we cannot have simultaneously a situation of maximum concreteness ($\sigma_q=0$) and maximum abstractness ($\sigma_p=0$), or situations where concreteness (resp., abstractness) would be maximal and abstractness (resp., concreteness) would be intermediary (i.e, with a finite standard deviation). However, the product $\sigma_q \sigma_p$ should be also bounded from above, as we cannot simultaneously have a situation of minimum concreteness ($\sigma_q=\infty$) and minimum abstractness ($\sigma_p=\infty$), or situations where concreteness (resp., abstractness) would be minimal and abstractness (resp., concreteness) would be intermediary (i.e., with a finite standard deviation). And in fact, a reverse version of Heisenberg's uncertainty relations can also be derived, as was recently done \cite{Mondaletal2017}, in accordance with what the conceptuality interpretation indicates.

\section{Explaining non-spatiality (non-locality)\label{non-spatiality}}

According to the above discussion, Heisenberg's (direct and reversed) uncertainty relations should not be considered to be the result of a lack of precision about how observables are measured in the laboratory, or the fact that measurements can alter the state of the measured entity (as was initially considered by Heisenberg in his semiclassical microscope reasoning). They would instead be an ontological statement describing the necessary tradeoff between concreteness and abstractness, resulting from the fact that, at the ontological level, quantum entities would be conceptual (meaning) entity. So, the \emph{non-locality} of a micro-entity like an electron, which should be more properly denoted \emph{non-spatiality}, would express the fact that most of the electron's states are abstract ones (with different degrees of abstractness), with the subset of the maximally concrete ones only corresponding to those describing specific localizations in space. Accordingly, the classical notion of \emph{object} (here understood as a spatiotemporal entity) corresponds to a conceptual entity that can remain for a sufficiently prolonged time in a maximally concrete state, which means that objects (classical entities) would just be limit cases of conceptual entities immersed in deterministic contexts that allow them to remain maximally concrete for a long time. 

\begin{figure}
\begin{center}
\includegraphics[width=14cm]{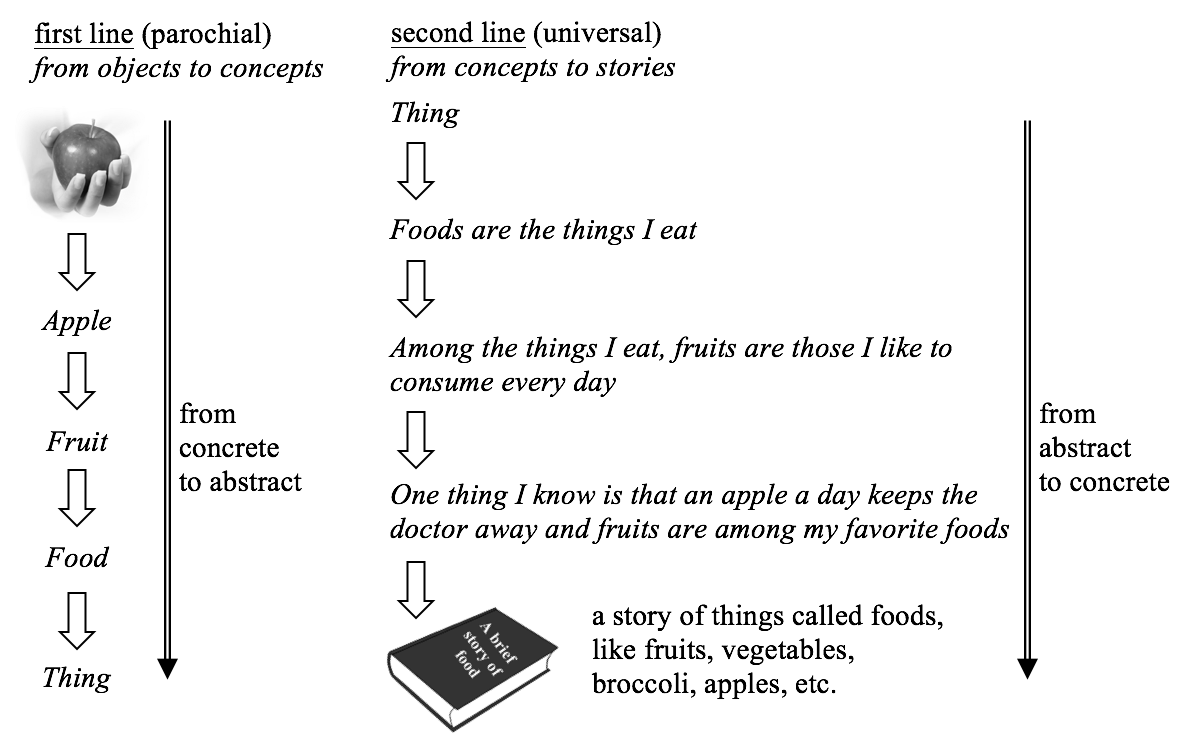}
\end{center}
\vspace{-15pt}
\caption{For human concepts there are two main lines connecting abstract to concrete. The first one goes from concrete to abstract: from objects to collections of objects having common features. The second one goes from abstract to concrete: from concepts to stories formed by the combination of many concepts.}
\label{Figure5}
\end{figure}

A possible criticism of the above explanation of Heisenberg's uncertainty relations would be that there is nothing truly fundamental in our human distinction between abstract and concrete concepts, as is clear that what we call concrete concepts are precisely those associated with the objects we have interacted with, in the course of our evolution on the surface of this beautiful ``pale blue dot.'' It is certainly true that physical objects have played an important role in the way we humans have formed our language and have created more abstract concepts, for instance when in the need of indicating an entire category of objects instead of just a member of a category. So, in this human historical line of going from the concrete to the abstract, the most concrete concepts are those specifying spatiotemporal entities (objects), like in the conceptual combination: \emph{This item that I'm presently holding in my hands}, and the most abstract ones are those indicated by terms like \emph{Entity}, \emph{Thing}, \emph{Stuff}, etc., with all the other concepts lying in between them, as regards their degree of abstractness/concreteness (see Figure~\ref{Figure5}). This human (parochial) line is the one typically considered in semiotics and psychology, which is the reason why psychologists use the term \emph{instantiation} to denote a more concrete form (a more concrete state) of a given concept. This term mostly refers to the actualization in time of an exemplar of a more abstract concept, like when \emph{Apple} is chosen as an exemplar of \emph{Fruit}, but of course one could also use the term \emph{spatialization} (or \emph{spatiotemporalization}), in addition to instantiation, when the exemplar in question is an object also existing in space. One should however bear in mind that a human concept, even when indicating an ordinary object, is not an object, and \emph{vice versa}, a physical object is not a human concept, although the latter can be put in a correspondence with the former and the former, according to the conceptuality interpretation, is a conceptual entity in a maximally concrete state. 

So, there is a parochial line to go from the concrete to the abstract, linked to the historical way we have developed concepts (starting from our need to name the physical entities around us), by abstracting them from objects, and there is a second line \cite{Aerts2014}, going from the abstract to the concrete, linked to how we humans have learned to combine concepts (in order to better think and communicate), creating more complex emergent meanings (see Figure~\ref{Figure5}). In this second line, the more abstract concepts are those that are expressible by single words, and concreteness increases when the number of conceptual combinations increases, so that the most concrete concepts are those typically described by large aggregates of meaning-connected (entangled) single-word concepts, which is what in our human realm we would generically indicate as \emph{stories}, like those written in books, articles, webpages, etc. We don't mean here stories only in the reductive sense of novels, but in the more general sense of clusters of concepts that are combined together in an interesting way, so as to create a well-defined meaning. This second line is therefore very different from the previous one (in a sense, it is transversal to it), and we humans clearly use both lines at the same time, when communicating and creating new meanings. However, it is this second line that we believe is the truly fundamental and universal one, i.e., the one in which the human concepts have found their natural developmental niche.

The fact that in human language both lines exist and are mixed together can explain in part the fact that there are structural differences between our human conceptual realm and the micro-physical conceptual realm, in particular the fact that the latter will generally exhibit a higher level of symmetry and organization (another reason being that our human cultural evolution is a recent happening relative to the time scale of our universe). Now, consider a document containing a text, and assume that the text contains the word ``horse.'' This means that the story in such document  is a (deterministic) context specifying a state of the concept \emph{Horse}, which according to the second line of concretization would be a very concrete state. Of course, this same document  can also be considered to describe the state of concepts indicated by other words in the story, or even concepts whose words are not specifically mentioned but are nevertheless strongly meaning-connected to its content. It is worth emphasizing that this document, containing a story about \emph{Horse}, is not necessarily associated with a physical horse that one can touch and ride (an instantiation of \emph{Horse}, according to the first line of concretization). For instance, the text may refer to the drawing of a horse, which of course is not a living entity, or maybe the term ``horse'' is only used in a metaphorical way, like in the Italian saying ``la superbia va a cavallo e torna a piedi'' (pride rides a horse and walks back). However, since the document  contains a whole story, the latter will behave in the conceptual realm in a way that is similar to how macro-physical objects also behave. 

To explain what we mean, consider two objects, let us call them object-A and object-B. When we consider the conceptual combination \emph{Object-A and object-B} (using the `and' logical connective),\footnote{To facilitate understanding, we will always denote concepts using italic type fonts and an uppercase first letter, to distinguish them from objects, which we will indicate using roman type fonts.} we are still able to put it in a correspondence with an object, and more precisely the object that is obtained by bringing together the two objects, now forming a single composite object (let us call it object-A$\wedge$B). On the other hand, when we consider the conceptual combination \emph{Object-A or object-B} (using the `or' logical connective), we are not able anymore to associate it with an object. But if we have two concepts, let us call them \emph{Concept-A} and \emph{Concept-B}, then not only \emph{Concept-A and concept-B} is again a concept, but also \emph{Concept-A or concept-B} is a concept. So, the conceptual realm is closed with respect to the conjunction and disjunction connectives, whereas the realm of objects is by definition only closed with respect to the conjunction connective (the conjunction of two objects is still an object, but the disjunction of two objects is not anymore an object).

What about stories, i.e., conceptual entities that are formed by large combinations of concepts that are connected through meaning? Of course, since a story is still a concept, and more precisely a concept that is obtained by consistently combining numerous other concepts, as described by the specific combination of words that are present in a document (like a book, a webpage, etc.) that makes the story manifest, the above must still hold: if we have two stories, let us call them \emph{Story-A} and \emph{Story-B}, then also \emph{Story-A and story-B} and \emph{Story-A or story-B} are to be considered stories. But the subtle point is in the distinction between the notion of story as a concept (i.e., a meaning entity) and the possibility for a story to be also manifest in concrete form in our spatial theater. Consider two actual books, let us call them book-A and book-B, with book-A containing the words of \emph{Story-A}, and book-B 
containing the words of \emph{Story-B}. What about the books associated with the two stories \emph{Story-A and story-B} and \emph{Story-A or story-B}? Let us call them book-A$\wedge$B and book-A$\vee$B, respectively. The former can simply be considered as the juxtaposition of book-A and book-B, which means that as soon as two books exist, each one telling a different story, then also the book containing the conjunction of their two stories can be considered to exist, and to correspond to the book obtained by placing the two books side by side. In other words, when looking at the shelves of a bookstore, with the different books placed in them side by side, we are actually contemplating stories that are conjunctions of other stories.

The situation is different when we consider the disjunction of two stories. In a bookstore, we will usually not find books of the form book-A$\vee$B. This not because it would be difficult to create an object of this kind, in our material world. Indeed, to do so, we only have to create a single page having the word ``or'' written on it, then consider the juxtaposition of book-A, such page, and book-B. But the probability to find an artifact of this kind in a bookstore is extremely low, and this because in our human culture it would not be considered to be the manifestation (the collapsed state) of a meaningful story, as is clear that for two arbitrary stories, \emph{Story-A} and \emph{Story-B}, the ambiguity introduced by the \emph{Or} connective will be considered to be too artificial for \emph{Story-A or story-B} to deserve to be engraved in a concrete document. To put it in a different way, in general the \emph{Or} connective in \emph{Story-A or story-B} will not provide a sufficiently strong meaning-connection for \emph{Story-A or story-B} to be able to also appear in a \emph{bona fide} book that humans can buy in bookstores. In other words, although in theory book-A$\vee$B, telling \emph{Story-A or story-B}, can be easily physically created, it will only appear with a very small probability within the field of our human cultural activity. The above does not mean, however, that stories that are disjunctions of other stories will not appear in documents that are part of our human culture. This will be the case of all texts that, for narrative reasons, require to specifically introduce such an aspect of two storylines that are told one after the other, with a disjunction in between. A typical example would be that of a detective story, in which different scenarios are told as possible solutions of a crime. Note that as we consider smaller pieces of texts, disjunctions will appear much more frequently, like in sentences of the ``coffee or tea'' and ``dead or alive'' kind. 

So, different from the disjunctions of stories, conjunctions of stories are in a much more obvious (concrete) way stories again, and this difference in behavior of stories in relation to the \emph{And} and \emph{Or} connectives indicates their special status as elements of greater concreteness of a conceptual realm. And in the same way as objects that are conjunction of other objects need more space to manifest in our spatial theater, stories that are conjunctions of other stories also need more ``space'' to manifest, i.e., more pages, more words, more memory on a computer, in case they would be electronic documents, etc. However, different from the ordinary objects, there is not yet for our human stories the equivalent of a well-structured spatial realm, and surely there are many different ways of defining the  embryonic structure from which a more organized and symmetric environment might one day emerge.

As a paradigmatic example, consider that specific collection of human stories that we have called the World Wide Web. Its interlinked webpages can be understood as the possible spatiotemporal manifestations of a rather complex abstract entity of meaning (formed by the combination of multiple concepts), whose full description requires the use of the quantum formalism (or even more general quantum-like formalisms). This perspective was recently considered in some detail as a way to capture the full meaning content of collections of documental entities, and the name QWeb was proposed to denote such meaning entity, to distinguish it from the spatiotemporal Web of written pages \cite{AertsEtal2017}. The QWeb, as a quantum-meaning entity, can be in different states: some of them will be more abstract, others more concrete, the most concrete ones being the stories associated with the different printable webpages. We can thus consider the entire collection of interlinked webpages as the equivalent of our three-dimensional Euclidean space, understood as a theater for those (classical) entities we call objects. In other words, we can consider the Web's collection of documents at a given moment of our human cultural history to be the equivalent (or rather, the embryonic version) of the possible spatial locations that micro and macro physical entities can occupy, be it in ephemeral or more permanent ways. 

This means that we interpret the different stories associated with the different webpages as the equivalent of the spatial states to which the QWeb entity (or some of its sub-conceptual entities) can transition to, in given experimental contexts, like for instance the interrogative context where a human inserts the word ``horse'' in the Google search engine, to obtain, as a result, a story about \emph{Horse}, among the different possible ones. Such a search experiment can be considered to be the equivalent of a quantum measurement, although of course the parallel is not complete, as is clear that search engines like Google still operate today in a deterministic way, whereas quantum measurements are genuinely indeterministic, as the decision processes operated by humans also probably are (see Sec.~\ref{measurement}). But we can certainly consider in our parallel a future versions of search engines, also integrating in their functioning probabilistic processes (i.e., some level of randomness), and in any case even today a human is always presented with a collection of possible results, ordered according to their relevance, and s/he has thus to decide on which of the obtained list of links to click, introducing in this way an element of unpredictability in the process.  

Before continuing with the discussion, let us stress again the double status of webpages: they have acquired the status of objects in our human world, as it is the case for all human artifacts, but they also describe complex conceptual combinations (what we have called stories) that correspond to the most concrete states of the QWeb conceptual entity. But not all human artifacts are necessarily associated with maximally concrete states of concepts, according the second line of concretization depicted in Figure~\ref{Figure5}. For instance, a piece of paper with the single word ``horse" written on it, is an entity in a maximally concrete state according to the first line (it is an object), but not an entity in a maximally concrete state  according to the second line (it is not a story).\footnote{In our Web analogy, we are assuming that humans are only motivated to create a webpage when it can convey a sufficiently articulated and complex meaning, and that a webpage containing, say, the single word ``horse," will not be deemed to be sufficiently interesting to justify the effort (the energy to be spent) for its creation, in the same way that we do not find on the shelves of a bookstore volumes whose pages, except for the cover title, would be all empty. But of course, artifacts of this kind are not in principle impossible to create, and in fact are also created. For instance, in a stationery shop, one can find notebooks, which are volumes without printed words. But a stationery is a very different context from that of library, or of a bookstore.} Having said that, we immediately see that a concept like \emph{Animal}, say in the state \emph{The animal is a horse}, which as we discussed can be considered to be a fairly abstract state, entertains a strong meaning-connection with a number of webpages, for instance all those containing the word ``horse,'' and this means that the conceptual entity \emph{Animal} is potentially present in all these webpages, i.e., in all these clusters of meanings that are stories about \emph{Horse} and which can be selected in an experiment consisting in finding a good example of a horse story. But since a conceptual entity can only be in a state at once, for as long as a webpage is not selected, we cannot say it is actually present in space (and time), as for this it has to acquire, at a given moment, one of the states belonging to the Web spatial canvas of states. 

We thus have here an interesting explanation of non-locality. First of all, as highlighted in many works even before the conceptuality interpretation was proposed, \emph{non-locality means non-spatiality} \cite{Aerts1998,Aerts1999}. Our three-dimensional Euclidean space (or more generally our four-dimensional Minkowskian space, possibly also curved by gravity) should not be considered the overall theater of our physical reality, but `a space' that emerged following the structuring of the macro-physical entities that grew out of the micro-ones. The conceptuality interpretation adds however an important piece of explanation, regarding how we should understand this notion of non-spatiality: \emph{non-spatiality means abstractness}. More precisely, non-locality and non-spatiality would result from the fact that the micro-physical entities being conceptual entities, stories (complex combinations of concepts) can form out of them, with \emph{coherence} (the structuring element for their formation) being nothing than the expression of a meaning-connection, exactly in the same way as, in the human conceptual realm, stories, and in particular webpages, originate and are structured through the meaning contained in individual and collective worldviews. And these meanings, connecting the more abstract concepts to the more concrete ones, explain why quantum conceptual entities are always available in acquiring spatial properties, by lending themselves to be detected by the physical apparatuses that belong to that semantic space (the Euclidean space) which is a theater for their stories. 

Consider a story mentioning an \emph{Animal} at different places in its narrative. Imagine that, at some moment, the \emph{Animal} gets specified as being a \emph{Horse}, i.e., the \emph{Animal} concept in the story enters \emph{The animal is a horse} state. Then, in no time, it will become a \emph{Horse} everywhere else in the story, where it was referred to as \emph{Animal}, which is precisely what happens in experiments when entities separated in space by large distances are observed to simultaneously change their states in a correlated way.

\section{Objects as limit of concepts\label{objects}}

It follows from the discussion above that what we usually call objects (classical entities having stable spatiotemporal properties) would be nothing but conceptual entities having reached the status of full-fledged stories, i.e., of sufficiently complex combinations of meaning-interconnected concepts. Again, we stress the importance of not confusing artifacts containing human stories (like printed webpages) with the fact that these human artifacts, as macroscopic material entities, are in turn story-like non-human conceptual entities. The notion of object, as used in classical physics, would then be only an idealization, as the object behavior would only depend on the conceptual/cognitive environment in which an entity is immersed. Consider the example of O'Connell mechanical resonator (a small 60 $\mu m$ flap, large enough to be seen with the naked eyes) which they succeeded putting in a superposition of two classically mutually exclusive states, one ``vibrating a little'' and the other ``vibrating a lot'' \cite{Connelletal2010}. As another example, consider the experiment performed by Gerlich et al., where organic molecules formed by up to 430 atoms, with maximum sizes of up to 60 angstrom, were successfully put in a superposition of states localized in regions of space separated by distances of orders of magnitude larger than the molecules' sizes \cite{Gerlichetal2011}. Experiments of this kind indicate that also big material entities, like chairs and tables, could in principle enter non-spatial states. Take a chair. If, at a fundamental level, it is also a story-conceptual entity, then it can be in different conceptual states. The most neutral one is simply the state expressing it existence, which we can describe by the conceptual combination \emph{The chair is a chair}, or \emph{The chair exists}. Other states of the chair-conceptual entity are easy to encounter in our human environment, like the state: \emph{The chair is in the bedroom} or \emph{The chair is in the livingroom}. These are eigenstates relative to the contexts where chairs are usually found. But in principle, and although this may well never be in our reach in experimental terms, one could also create interrogative contexts, like those considered by Gerlich et al. for the organic molecules, where a chair's state would be described for instance by the conceptual combination: \emph{The chair is in the bedroom or in the livingroom}. 

The enormous difficulty in obtaining in practice a state of this kind is due to the fact that a chair is a very complex object, i.e., a very complex story, formed by numerous sub-stories, and that to put an entity of this level of complexity in a state of superposition is about finding a way to decompose such story into what can be described as the disjunction of two different stories: one corresponding to the chair being in the spatial state \emph{The chair is in the bedroom}, and the other one corresponding to that same chair being in the spatial state \emph{The chair is in the livingroom}.\footnote{The conceptuality language is very fluid: a conceptual combination used to describe the state of a conceptual entity can also in turn, depending on the context considered, be interpreted as a composite entity of its own. Here the focus is on the entity \emph{Chair}, so a combination like \emph{The chair is in the bedroom} is to be interpreted as a specification of one of its possible states, but \emph{The chair is in the bedroom}, as a combination of 6 different concepts, can also be interpreted as a multipartite conceptual entity, which in turn can also be in different states.} So, to obtain a state like \emph{The chair is in the bedroom or in the livingroom}, for a macroscopic material entity like a chair, an experimental setting playing the role of a mind-like cognitive entity needs to be put in place, able to consistently decompose its meaning in a way that we would precisely describe as the disjunction of two different chair-stories (without destroying the chair-entity). Note that human minds can easily create such an interrogative context, when they express a lack of knowledge about where the chair actually is, and formulate such uncertainty-ambiguity situation using the ``or'' connective, i.e., producing a more abstract state.\footnote{To conveniently describe the conjunction as a superposition state, the lack of knowledge in question needs to be a deep one, such that one does not even know if the chair is either in the bedroom or in the living room, i.e., if the chair is or not in a spatial state. In other words (see the analysis of the double-slit and delayed choice experiments in Secs.~\ref{double}-\ref{delayed}), the ``or'' needs to be understood in an non-exclusive way.} This means that within the human conceptual realm, human minds can easily provide a context/interface that can interact with a chair-entity in a conceptual way, i.e., put it in a superposition state that they can subsequently collapse, when some additional knowledge is acquired. This should not be misinterpreted, however, as human minds objectively collapsing the physical chair, as considered in `consciousness causes collapse' interpretations  like the von Neumann-Wigner.  Again, we have not to confuse human concepts with the conceptuality of the physical entities, and human cognition with the cognitive behavior of the material entities (like the measuring apparatuses) that are sensitive to the meaning carried by the conceptual physical entities. 

So, can we create a physical context able to put a chair in a superposition state, corresponding to two different locations,\footnote{In other words, a superposition that is experienced as such by all the material entities playing the role of minds with respect to the proto-language of which the chair would be part of, and not a superposition for the human minds experiencing a doubt regarding the location of the chair, and expressing it by means of the disjunction connective.}  and at the same time also provide an interface able to conceptually interact with the chair in such a superposition state, i.e., to understand the meaning it carries and possibly subsequently collapse it onto states having well-defined spatial properties, as we can do with microscopic and mesoscopic physical entities? As we said, our tentative answer is that this should in principle be possible, and the fact that so far we have idealized entities like chairs as objects, instead of conceptual entities, is because their conceptual nature can only manifest when a context of the double-slit kind is created for them. But what could be considered the equivalent of a detecting screen for an entity like a chair? We can observe that since our standard terrestrial environment is able to maintain macroscopic bodies constantly in space, then this same environment can be expected to be able of also producing the collapse -- the objectification -- of a macro-entity like a chair in a superposition state. But then, how can we bring an entire chair in a more abstract state, of spatial superposition? Why would it be so difficult to do so, in comparison to, say, a hydrogen atom? The answer is simple: for larger and larger entities, it becomes increasingly difficult to obtain an effective shielding from the unceasing random thermal bombardment to which they are subjected to, on the surface of our planet, and there is not only the external bombardment: also the internal environment of the chair needs to be taken into account. 

To explain what we mean, we can reason as follows. To put the whole chair in a superposition of two different spatial locations, we have to be able to describe the chair-entity as a coherent whole. In mathematical terms, this can be translated into the possibility of factorizing the wave function in a way that the center of mass contribution separates from the contribution coming from the movements of the different constituents, relative to that center and to each other's. Indeed, it is not the part of the wave function describing the relative motion of the internal components that we want to put in a superposition state (as this part of the wave function describes the structure of the chair, which we want to preserve), but that describing its center of mass, which describes the potential localization in space of the chair. In the case of a hydrogen atom, it is straightforward to separate the wave function relative to the center of mass from that associated with the relative motion, obtaining in this way a description of the evolution of the center of mass as a free evolving wave function (see any manual of quantum mechanics). But with a macroscopic body things get much more complicated, as to be able to describe the chair's center of mass by means of a free evolving wave-packet, the evolution of the body's center of mass needs to decouple from all internal degrees of freedom, and this can reasonably be done only if the body is cooled down to extremely low temperatures. How low? Well, low enough to avoid any exchange of energy between the center of mass degree of freedom and the degrees of freedom associated with all the internal relative movements~\cite{Sun2001}.

One may wonder why these exchanges of energy would be so problematic. It is of course easy to understand that the external bombardment of heat packets of energy can cause what is usually denoted as \emph{loss of quantum coherence}, which within the conceptuality interpretation translates into \emph{loss of meaning}. This loss of meaning is caused by the fact that when a physical system is forced to communicate with a noisy environment, this will consequently blurr also the internal communications, with the result that the internal components will cease to behave as a coherent whole. But even if the external bombardment would not be thermal, but fully coherent, this would probably not solve the problem of the blurring of the internal communications of the chair-entity. Indeed, a chair is a very complex entity, made of innumerable parts, some of which are more cohesive than others. It is like an environment formed by different individuals, with different brains, so that even when they all receive the same input, like a spoken sentence (a concept in a given state), this will trigger a response that will differ depending on the individual involved. And of course, if numerous individuals are forced to chat together, all at the same time, without any coordination, producing each of them a different output, the overall result will be an unintelligible cacophony. This is what we can expect to happen in a photon-mediated communication happening at the level of the different pieces of matter that form the stuff the chair is made, like atoms, molecules, macromolecules, and other more or less separated coherent domains, because of the incessant processes of excitation and de-excitation. 

The problem of this discordant and meaningless mixture of different communications can in principle be solved by silencing all the participants, taking away their energy by cooling down the chair-entity to extremely low temperatures, and of course do the same with its external environment. In these conditions of extremely cold external and internal environments, also a chair, at least in principle, could be brought into a non-spatial superposition state, by letting it interact with a macro equivalent of a double-slit context. Now, considering once more the parallel between a complex entity like a chair and the notion of story, we can observe that also in our human written stories there are parts of them that are more cohesive than others. Take the example of a novel: different chapters, which are like sub-stories, can be distinguished, and then there are the paragraphs, usually also containing more cohesive and self-contained ``units of discourse,'' associated with given ideas (so that each paragraph can be considered to be a conceptual entity in its own, in the specific state described by the combinations of words in the paragraph). But going even further, and considering the more specific human line of concretization, we can also observe in human documents the presence of the ``and'' and ``or'' connectives, with the former being usually much more abundant than the latter. As we observed already, the connective ``or'' usually increases the level of abstraction, whereas the connective ``and'' would typically go in the direction of making the combination more concrete. Clearly, \emph{Duck and mole} conveys a much more specific meaning than \emph{Duck or mole}, as the ``or'' is more easily associated with a new possible emergent meaning, not reducible to those conveyed by \emph{Duck} and \emph{Mole} taken separately, and which in the long run might acquire a brand new term for its designation.\footnote{The effects of the ``and'' and ``or'' conjunctions as regards making a combination more or less abstract is in fact much more articulated; see for instance the discussions in \cite{Aerts2010b,Aerts2014}.} 

This breaking of symmetry between the ``and'' and ``or'' in human documental entities, and the fact that, generally speaking, the ``or'' connective produces stronger connections in meaning than the ``and'' connective (compare for instance \emph{Dead or alive} with \emph{Dead and alive}, \emph{Trick or treat} with \emph{Trick and treat}, etc.), is indicative of the fact that different \emph{domains of meaning} exist within texts, where the concepts belonging to these domains are much more submerged in each other’s meaning, so that a clustering of documents in meaning structures of different sizes is inherent in the way a meaning type of interaction works at a fundamental level. And of course, the clustering process causes an objectification process, with the larger clusters usually attaining a stronger object status within the governing meaning type of interaction. And in a physical object like a chair, the same will happen, if we understand quantum coherence to be the equivalent of meaning in the case of micro-entities: there will be domains of coherence within a chair, separated from other \emph{domains of coherence}, which in fact makes a chair, with good approximation, an entity formed by the conjunction of different parts with almost no meaning connection between them (no superposition), i.e., behaving almost as different objects. However, their conceptual nature can still be revealed, if an appropriate experimental context is put in place, like a context where the overall energy is lowered to a point where the \emph{de Broglie wave length} associated to all these separated domains can overlap and start to intimately communicate (for a detailed discussion of the notion of de Broglie wave length, see \cite{Aerts2014}). Let us mention here \emph{en passant} the difference between a dead piece of matter, like a chair, and a living piece of matter, like a platypus. One can say that the latter, different from the former, was able to construct, at room temperature, structures with nested domains of coherence (meaning) of all possible sizes, up to the size of the entire body of the living entity.

\section{Entanglement\label{entanglement}}

After what we discussed already in the previous sections, it becomes more easy to explain how entanglement can be accounted for in a satisfying way in the conceptuality interpretation. Entanglement is among the better studied and experimentally verified quantum phenomena, and one that appears to defy our common (spatial) sense, which is the reason why Einstein famously described entanglement as a ``spooky action at a distance.'' Indeed, the possibility of creating a condition of entanglement between two micro-entities appears not to depend on the spatial distance separating them or, to put it in more precise terms, appears not to depend on the spatial distance between the locations where the entangled entities can be detected with high probability. A characteristic of quantum entanglement (a direct consequence of the superposition principle) is that it is ubiquitous,\footnote{This explains why standard quantum mechanics cannot consistently describe separated physical systems \cite{Aerts1984}; see also Sec.~3 of \cite{Aerts2014}.} in the sense that quantum entities naturally entangle whenever they are allowed to interact and will typically remain entangled for as long as nothing intervenes to disentangle them (to decohere them). This ubiquitousness of entanglement mirrors the ubiquitousness of the \emph{meaning-connections} that are unavoidably present in any conceptual realm. As soon as two conceptual entities are allowed to meet in a given cognitive context, a meaning-connection will exist between them, whose strength will of course depend on how much meaning the two entities can share and exchange. 

Take the example of the two concepts \emph{Animal} and \emph{Acts}. These are two abstract concepts that are quite strongly meaning-connected in most contexts, as we all know from our experience of the world that animals are living beings and that living beings can do different types of actions, and that there are actions that certain animals will typically do that other animals will not do. This connection becomes perfectly evident when these two concepts are combined in a sentence like \emph{The animal acts}. Almost all human minds will agree that such sentence possesses a full and perfectly understandable meaning. To better understand the nature of this meaning-connection between \emph{Animal} and \emph{Acts}, when combined as \emph{The animal acts}, we can consider two couples of exemplars for both concepts, like the following ones \cite{as2011}: (\emph{Horse}, \emph{Bear}) and (\emph{Tiger}, \emph{Cat}) for \emph{Animal}, and let us denote these two couples $A$ and $A'$, respectively, then (\emph{Growls}, \emph{Whinnies}) and (\emph{Snorts}, \emph{Meows}), for \emph{Acts}, and let us denote them $B$ and $B'$, respectively. One can then invite a certain number of individuals to participate in the following coincidence experiment. Considering the combination \emph{The animal acts}, they are asked to select pairs of exemplars for these two concepts, as representative examples of their combination. If they choose from the couples $A$ and $B$, their selection will be considered to be the outcome of a joint measurement denoted $AB$, and similarly for the other combinations, thus defining in total four joint measurements: $AB$, $A'B$, $AB'$ and $A'B'$. The statistics of all these outcomes can then be analyzed in the same way physicists analyze data of Bell-test experiments, for instance using that version of Bell's inequality known as the Clauser, Horne, Shimony and Holt (CHSH) inequality \cite{clauser1969}.\footnote{The CHSH inequality is $|S|\le 2$, with $S\equiv E(A,B)-E(A,B')+E(A',B)+E(A',B')$, where $E(A,B)$ denotes the expectation value for the joint measurement $AB$, given by: $E(A,B)=p(A_1,B_1)-p(A_1,B_2)-p(A_2,B_1)+p(A_2,B_2)$, with $p(A_1,B_1)$ the probability for obtaining the pair of outcomes $(A_1, B_1)$, i.e., (\emph{Horse}, \emph{Growls}), corresponding to the combination \emph{The horse growls}, and similarly for the other probabilities and joint measurements.} And the result is that the inequality will be violated with magnitudes similar to those of typical laboratory physics' situations with entangled spins or entangled photons \cite{as2011,as2014}.

So, the conceptual combination \emph{The Animal Acts} describes what in physics is considered to be an entangled state. Such combination contains both a specification of the state of \emph{Animal} and of the state of \emph{Acts}, but also a specification of the ``state of their connection.'' Indeed, if instead of \emph{The animal acts} we would have used the more complex combination \emph{An animal that has been doped acts in a strange way}, not only the specification of the states of \emph{Animal} and \emph{Acts} would be different, but also their meaning connection, in that context, would be different. This however is not how one would usually interpret an entangled state in quantum mechanics. Indeed, since genuine quantum states are only assumed to be described by pure states, and one cannot associate pure states to the different components of a composite entity when they are entangled, the usual conclusion is that when a composite entity is in an entangled state, its components cease to exist, in the same way as two water droplets also cease to exist when they are fused into a single larger droplet. This however is not fully consistent with the observation that entanglement preserves the structure of the composite entity. For instance, two entangled electrons, when disentangled, will still have the same mass and electric charge. In other words, quantum entities certainly do not completely disappear when entangled, as the conceptual combination \emph{The animal acts}, interpreted as an entangled states, also indicates. So, do we have an incompatibility between the conceptuality interpretation and what the standard quantum formalism indicates? Surely not, though the conceptuality interpretation certainly pushes us towards a completion of the quantum formalism, to also allow the components of an entangled system to remain in well-defined states. This can be done by adopting the recently derived \emph{extended Bloch representation} (EBR) of quantum mechanics \cite{AertsSassoli2014}, where one can consistently represent the state of a bipartite composite entity as a triple of (real) vectors, with two of them specifying the individual states of the two components and the third one (of higher dimensionality) the state of their connection \cite{AertsSassoli2016a,AertsSassoli2016b}. The reason why the extended Bloch representation can do so is that it is a completed version of the quantum formalism, where density operators also play a role as representatives of genuine states, so that one no longer needs to give up the general principle saying that a composite system exists, and therefore is in a well-defined state, if and only if its components also exist, and therefore are in well-defined states (see also Sec.~\ref{measurement}, for the role played by the EBR in relation to the measurement problem).

As soon as we explain entanglement as a meaning-connection, the phenomen is demystified. First of all, because it becomes clear that there are no communications through space that should be associated with the quantum correlations, as a meaning-connection between two concepts is an abstract element of reality, not manifesting at the level of our spatial theater. And it is also clear that although it is correct to describe an entangled system, like two entangled electrons, as a whole, because of the presence of the meaning-connection playing the role of a connective element, not for this one should think that the conceptual entities would have lost their identity in the combination (in a nutshell, entanglement, as an emergent phenomenon, is not ``$1+1\to 1$,'' but ``$1+1\to 3$''). And once one considers the role played by this connecting element, it becomes evident that when individual properties are created (instead of just discovered) in a coincidence measurement, also correlations will be created (instead of just discovered), precisely because of the presence of a non-spatial (more abstract) connection. In other words, it is because correlations are created in a joint measurement (called \emph{correlations of the second kind} \cite{Aerts1991,AertsSassoli2016b}), instead of just discovered, that Bell's inequalities can be violated, and the only way to create correlations out of a composite entity is to have the components to be connected prior to the measurement. 

\begin{figure}
\begin{center}
\includegraphics[width=10cm]{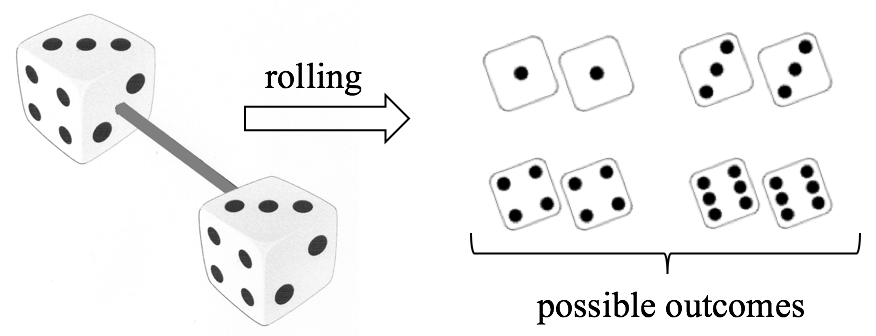}
\end{center}
\vspace{-15pt}
\caption{When two disconnected dice are jointly rolled, there are 36 equiprobable pairs of upper face-outcomes (no correlations). But if the two dice are connected through a rigid rod, when they are rolled only 4 pairs of upper face-outcomes can be obtained (correlations are created by the rolling experiment.}
\label{Figure6}
\end{figure}

To help better understand what we mean by this, consider two traditional dice. If we roll them at the same time, we will obtain 36 possible and equiprobable pairs of upper face-outcomes. This is a situation where no correlations can be detected in the statistics of outcomes. However, if we connect the dice through a rigid rod, then only certain couples of upper faces can be obtained, when they are jointly rolled, and not others (see Figure~\ref{Figure6}). In this example, we can perfectly see the role played by the connecting element, here perfectly visible as a connection through space, and composite interconnected macro-systems of this kind can easily violate Bell's inequalities \cite{sdb2013}. 
The connection through meaning plays exactly the same role of the rigid rod connecting the two dice through space. Of course, it will not work in such a stable way, when human minds select couples of exemplars representative of more abstract concepts, as fluctuations are also expected to be present. Remaining within the paradigm of the dice example, a more realistic description would be that of a rigid rod having a probability of also detaching and falling during the execution of the joint rolling process, so that correlations will not be always perfect, which is something that will typically lower the degree of violation of Bell's inequalities \cite{sdb2014}. 

To complete our discussion, let us also give the example of a conceptual situation that would be the equivalent of two micro-entities in a non-entangled (product) state, like two disconnected dice. Consider the conceptual combination \emph{The animal is a cat whose favorite act is to meow}. Since such combination already actualizes a connection between \emph{Cat} and \emph{Meow}, the process of creating correlations during the joint measurement will be considerably reduced. In other words, we are here in a situation of \emph{correlations of the first kind} \cite{Aerts1991,AertsSassoli2016b}, which will be typically discovered instead of created during the experiment. More examples could be provided of human conceptual situations mimicking what happens in entangled micro-system, when interpreted as conceptual meaning-connected systems. A quite suggestive cognitive psychology experiment was for instance recently performed, where participants were asked to select pairs of wind directions they considered to be good representatives of \emph{Two different wind directions}, with the data showing a violation of the CHSH inequality of magnitude close to that of experiments with entangled spins \cite{aabgssv2017a}. A symmetrized version of the experiment was also considered, which received a complete quantum modeling in Hilbert space, using a singlet state to describe the meaning-connection and product measurements to describe the interrogative context where couples of actual wind directions were selected \cite{aabgssv2017b}. 

Let us use this last example of wind directions to make it even more explicit the parallel between the nature and behavior of conceptual and micro-physical entities. When we consider the conceptual combination \emph{Two different wind directions}, none of the two winds concepts in it has a spatial direction. In the same way, considering two spin-${1\over 2}$ quantum entity in a singlet (entangled) state, also in this situation the two spins have no spatial direction. These are only acquired when the two spin entities are forced by the measuring apparatus to acquire one, in the same way that participants to the cognitive experiment are forced to choose actual couples of wind directions. The way they do so depends on the accumulated human experience with winds blowing on the surface of our planet, and the meaning that was abstracted from these experiences. This will cause certain wind directions to be perceived as more different than others, thus favoring a process of creation of strong correlations during the selection of pairs of spatial directions that are judged be the best examples of \emph{Two different wind directions}. This is exactly what also happens in coincidence experiments with two spin-${1\over 2}$ entities in a singlet state, which is a state of zero spin where specific directions (eigenstates of the spin operators) have not yet been created. So, when the mind-like Stern-Gerlach apparatuses jointly select a spin direction, i.e., when they answer the question ``What is the best example of two different spin directions?'' they will produce an answer taking into account the meaning content carried by the composite conceptual spin entity, which can be described by the conceptual combination \emph{The total spin value is zero} or, to express it in even more specific terms: \emph{Spin orientations are always opposite when they are created along a same direction}.

\section{Indistinguishability}
\label{indistinguishability}

In the previous section, we explained that entanglement, according to the conceptuality interpretation, is the expression of a meaning-connection between conceptual entities. Sometimes, entanglement is described as \emph{quantum coherence}, where the term ``coherence'' is to be understood as a given, fixed relation between states, which is precisely what an entangled state is: a fixed relation between product states expressed by means of their superposition. This relation, or connection, is a meaning-connection existing before the entangled entities are subjected to possible interrogative contexts. So, realism is clearly not at stake when dealing with entanglement, as reality, as we explained already, would not be fully contained in the spatiotemporal theater and entangled quantum entities would be entities in more abstract states, available to acquire spatial properties (like locations and directions) only when submitted to suitable contexts, like the measurement ones. In other words, we have to distinguish what connects entities, and the effects that these connections produce in terms of correlations that can be created in the laboratories, which are processes where more concrete exemplars/instantiations of abstract concepts can be jointly actualized. 

In this section, we want to address another of the quantum conundrums, \emph{indistinguishability}, and explain why it can be convincingly elucidated by the conceptuality interpretation; this because concepts have a built-in notion of indistinguishability, which is apparently what we also use by default when we deal with large collections of concepts \cite{Aerts2009a, Aertsetal2015b}. But before that, let us briefly recall what the notion of indistinguishability is about. Two entities -- let us call them $S_1$ and $S_2$ -- are said to be distinguishable if when we exchange their role this can have observable effects, at least in principle. Entities that are indistinguishable are said to be \emph{identical}, and identical means that they possess exactly the same set of attributes, i.e., the same set of state-independent intrinsic properties, like for instance a same mass, charge and spin, as it is the case for all elementary micro-entities, for example electrons. Now, identical entities, although indistinguishable, are nevertheless \emph{individuals}. This is precisely because they have attributes that can be measured and used to count how many of them are present in a composite system. For instance, the electric charge of a collection of electrons, if measured, will be $Ne$, with $e$ the charge of a single electron and $N$ an integer number indicating the number of identical electrons that are present in the collection. Hence, identical individuals are not necessarily a same individual, i.e., what renders two entities distinguishable appears not to be what also confers them their individuality. 

\begin{figure}
\begin{center}
\includegraphics[width=14cm]{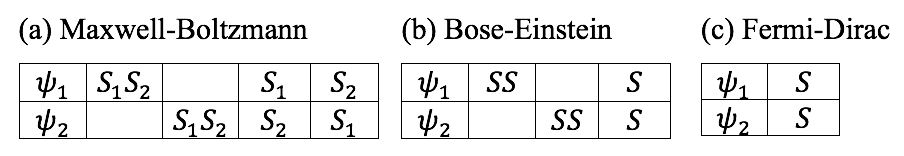}
\end{center}
\vspace{-15pt}
\caption{The total number of states for two entities that can be in two different states, $\psi_1$ and $\psi_2$, when (a) they are distinguishable (spatial objects); (b) they are indistinguishable and can be in the same state (bosons); (c) they are indistinguishable but cannot be in the same state (fermions).}
\label{Figure7}
\end{figure}

Indistinguishability has profound consequences on the statistical behavior of identical entities, when considered collectively. Consider first the case where $S_1$ would be in some way distinguishable from $S_2$, and assume for simplicity that they can only be in two different states, let us call them $\psi_1$ and $\psi_2$. Then, the two entities, when considered as a composite system formed by two non-interacting sub-entities, can be in 4 different states [see Figure~\ref{Figure7}(a)]: one where both entities are in state $\psi_1$, one 
where both entities are in state $\psi_2$, one where $S_1$ is in state $\psi_1$ and $S_2$ is in state $\psi_2$, and finally one where $S_1$ is in state $\psi_2$ and $S_2$ is in state $\psi_2$. In the more general case where the number of distinguishable entities is $n$ and the number of states they can be in is $m$, it is not difficult to see that the total number $N$ of states of the composite system formed by $n$ non-interacting sub-entities is: $N_{\rm MB}=m^n$, where the subscript ``MB'' stands for ``Maxwell-Boltzmann,'' as this way of counting is characteristic for the classical \emph{Maxwell-Boltzmann statistics}. 

But consider now the situation where the two entities are indistinguishable. In this case, the situation where $S_1$ is in state $\psi_1$ and $S_2$ is in state $\psi_2$, and the situation where $S_1$ is in state $\psi_2$ and $S_2$ is in state $\psi_1$, cannot anymore be distinguished, hence they correspond to the same situation, which means that now we only have a total of 3 different states [see Figure~\ref{Figure7}(b)]. Again, a formula can be written for the general case: $N_{\rm BE}= \binom{n+m-1}{n} = \frac{(n+m-1)!}{n!(m-1)!}$, where the subscript ``BE'' stands for ``Bose-Einstein,'' as this way of counting is characteristic for the quantum \emph{Bose-Einstein statistics}. For completeness, let us also describe a third situation, where not only the two entities are indistinguishable, but there is also the constraint that they cannot be jointly in the same state (Pauli's exclusion principle). Then only a single state remains for the composite system [see Figure~\ref{Figure7}(c)], and for the general situation we have the formula: $N_{\rm FD}= \binom{m}{n} = \frac{m!}{n!(m-n)!}$, where the subscript ``FD'' stands for ``Fermi-Dirac,'' as this way of counting is characteristic for the quantum \emph{Fermi-Dirac statistics}.

If the conceptuality interpretation correctly captures the nature of quantum entities, then quantum indistinguishability should appear also in the ambit of the 
human conceptual realm, at least to some extent, and produce non-classical statistics, not deducible from the MB way of counting states. Let us take the example of the abstract concept $\emph{Animal}$. We can consider a certain number of these $\emph{Animal}$ concepts, say ten of them. A collection of this kind can be described by considering the two-concept combination: \emph{Ten animals}. It is then clear that all the \emph{Animal} concepts in the combination are completely identical and all exactly in the same state, i.e., all carrying exactly the same meaning, and that we are truly in the presence of a collection of entities, not of a single one. In other words, in the conceptual combination \emph{Ten animals} the quantum indistinguishability becomes perfectly self-evident, so that the conceptuality interpretation offers a very simple and clear explanation of it. This would not possible for spatial objects, as is clear that two spatial objects are never indistinguishable, as they always occupy different locations in space, i.e., they are always in different spatial states. They can in principle all have the same intrinsic properties, but because of their spatiotemporal status they will always be distinguishable. So, the fact that \emph{Ten animals} is a concept, and not an object, is crucial for it being able to carry the quantum feature of `being many and at the same time being genuinely indistinguishable'.

Let us consider then two possible exemplars of \emph{Animal}: \emph{Cat} and \emph{Dog}. These are to be considered as two possible states of \emph{Animal}, i.e., the states expressing the meaning that \emph{The animal is a cat} and \emph{The animal is a dog}, respectively. We are thus in the situation where $m=2$ and $n=10$, so that $N_{\rm BE}=11$. More specifically, the eleven states that the concept \emph{Ten animals} can be in, when only the two exemplars \emph{Cat} and \emph{Dog} are considered, are: $\psi_{10,0}=$ \emph{Ten cats}, $\psi_{9,1}=$ \emph{Nine cats and one dog}, $\psi_{8,2}=$ \emph{Eight cats and two dogs}, \dots , $\psi_{2,8}=$ \emph{Two cats and eight dogs}, $\psi_{1,9}=$ \emph{One cat and nine dogs} and $\psi_{0,10}=$ \emph{Ten dogs}. If we assume that the \emph{Cat} and \emph{Dog} states can be actualized with the same probability and that there are no ways to distinguish between the individual cats, nor between the individual dogs, then the probabilities for obtaining all these states are the same, and given by $P_{\rm BE}(\psi_{10-i,i})={1\over 11}$, $i=0,\dots,10$. On the other hand, in case there would be an underlying reality allowing to make further distinctions, then all these states would have a multiplicity. More precisely, the multiplicity of the state $\psi_{10-i,i}$ is ${10!\over i!(10-i)!}$, which gives the MB probabilities: $P_{\rm MB}(\psi_{10-i,i})={10!\over i!(10-i)!2^{10}}$, $i=0,\dots,10$. More specifically: $P_{\rm MB}(\psi_{10,0})=P_{\rm MB}(\psi_{0,10})={1\over 1024}$, $P_{\rm MB}(\psi_{9,1})=P_{\rm MB}(\psi_{1,9})={5\over 512}$, $P_{\rm MB}(\psi_{8,2})=P_{\rm MB}(\psi_{2,8})={45\over 1025}$, $P_{\rm MB}(\psi_{7,3})=P_{\rm MB}(\psi_{3,7})={15\over 128}$, $P_{\rm MB}(\psi_{6,4})=P_{\rm MB}(\psi_{4,6})={105\over 512}$, $P_{\rm MB}(\psi_{5,5})={63\over 256}$.

\begin{figure}
\begin{center}
\includegraphics[width=12cm]{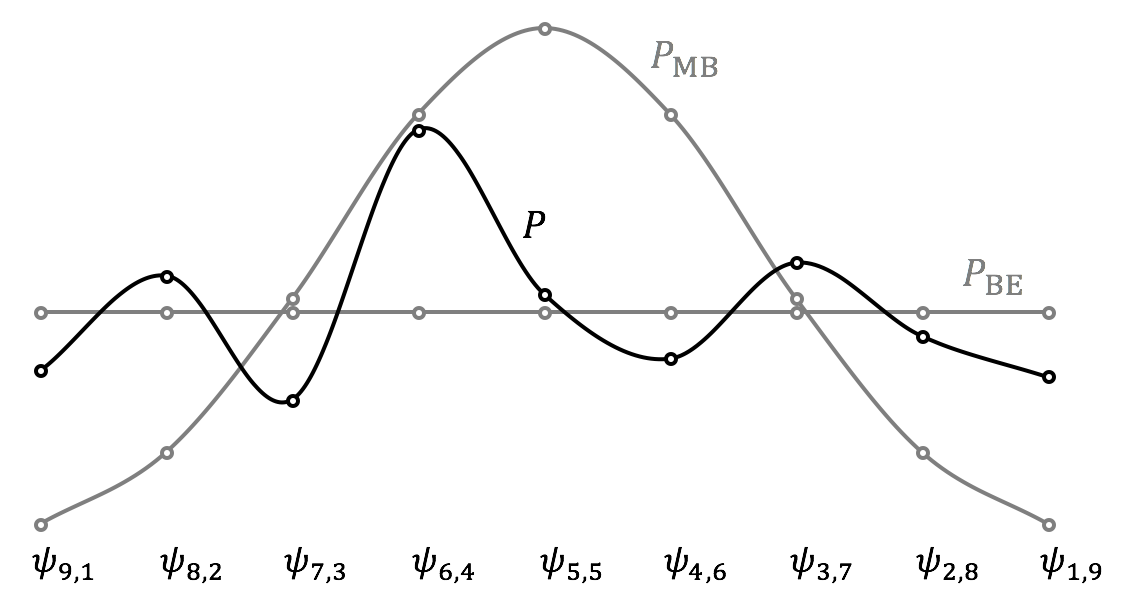}
\end{center}
\vspace{-15pt}
\caption{A comparison of the Maxwell-Boltzmann ($P_{\rm MB}$) and BE ($P_{\rm BE}$) probabilities with those obtained by performing Google's counts (on August 18, 2017) on the Web ($P$), in the situation where the conceptual entitiy \emph{Ten animals} is considered in relation to the two exemplar-states \emph{Cat} and \emph{Dog}. Note that the extremal states \emph{Ten cats} and \emph{Ten dogs} were not considered in the calculation.} 
\label{Figure8}
\end{figure}

Can we find evidence for a deviation from the MB statistics to the BE one, due to the indistinguishability of the individual $\emph{Animal}$ concepts in the combination \emph{Ten animals}? A possibility is to view the Web as a mind-like entity that can tell different stories, associated with all its searchable webpages. In this way, one can perform counts, using a search engine like Google, and use the obtained numbers as an estimate of the different probabilities (see \cite{AertsEtal2017} for more details about this way of interrogating the Web). When doing so, however, it is important to exclude the two extremal states $\psi_{10,0}=$ \emph{Ten cats} and $\psi_{0,10}=$ \emph{Ten dogs}, as these combinations will obtain counts that are two orders of magnitude greater than all the others, and this because the sentence ``ten cats'' (resp., ``ten dogs'') does not contain the ``dog'' (resp., ``cat'') word, and can thus easily combine with all possible other words. Furthermore, if we would use the more specific combination ``ten cats and zero dogs'' (resp., ``ten dogs and zero cats''), we would obtain no counts, as we don't usually express things in this way in conventional human language. Thus, our Web interrogation will not provide correct data for the two states $\psi_{10,0}$ and $\psi_{0,10}$, which therefore must be dropped from the statistics. This means that we only start counting the number of pages containing the combinations ``nine cats and one dog'' or ``one dog and nine cats.'' On August 20, 2017, Google gives: $N_{9,1}=3090$. Doing the same for the combinations ``eight cats two dogs'' or ``two dogs and eight cats,'' we obtain: $N_{8,2}=4790$, and proceeding in the same way, we find: $N_{7,3}=2580$, $N_{6,4}=7390$, $N_{5,5}=4460$, $N_{4,6}=3310$, $N_{3,7}=5020$, $N_{2,8}=3710$,  $N_{1,9}=2980$. With $N=N_{9,1}+N_{8,2}+\cdots + N_{1,9}=37330$, we can thus calculate the weights $P(\psi_{10-i,i})={N(10-i,i)\over N}$, $i=1,\dots,9$, and interpret them as the experimental probabilities for the states $\psi_{10-i,i}$, $i=1,\dots,9$. These are: $P(\psi_{9,1})= 0.083$, $P(\psi_{8,2})= 0.128$, $P(\psi_{7,3})= 0.069$, $P(\psi_{6,4})= 0.198$, $P(\psi_{5,5})= 0.119$, $P(\psi_{4,6})= 0.089$, $P(\psi_{3,7})= 0.134$, $P(\psi_{2,8})= 0.099$ and $P(\psi_{1,9})= 0.080$. In Figure~\ref{Figure8}, we represent them together with the theoretical MB ($P_{\rm MB}$) and BE ($P_{\rm BE}$) probabilities (after having renormalized them, following the cut off of the extremal states). Clearly, the data obtained from the counts  are much more typical of a BE statistics, with some added fluctuations, than a classical MB one.\footnote{Note that these fluctuations are really such, in the sense that the deviations from the Bose-Einstein ``flat line'' will be generally different when different concepts are considered, say for instance \emph{Horse} and \emph{Cow} instead of \emph{Cat} and \emph{Dog}, to stay on animals. In other words, the observed deviations from the Bose-Einstein statistics cannot be generally attributed to a systematic classical multiplicity of the states.}

Of course, Google's counts are far from being a precise estimate of the actual number of existing webpages containing specific combination of words, which means that the above is to be only considered as an illustrative example, more than a demonstrative one. More examples of Web counts can be found in \cite{Aerts2009a, Aertsetal2015b}. But more importantly, in \cite{Aertsetal2015b} experiments on human subjects were also performed. More precisely, 88 participants were given a list of concepts, like \emph{Eleven animals}, \emph{Nine humans}, \emph{Eight expressions of emotion}, etc., in association with two of their possible exemplars, like \emph{Cat} and \emph{Dog} for \emph{Animal}, \emph{Man} and \emph{Woman} for \emph{Human}, \emph{Laugh} and \emph{Cry} for \emph{Expression of emotion}, etc. More precisely, different numerical combinations of exemplars were each time presented to them, for each one of the concepts, asking them to evaluate what are the most probable combinations, according to their preference. The obtained result show that the passage from the BE statistics (corresponding to a perception of strict indistinguishability of the concepts) to a classical MB one, depends on the concepts and exemplars considered in the experiment, in the sense that the easier it is to relate them to everyday life situations, and the more the obtained statistics will tend towards the MB one. On the other hand, the less the human imagination is influenced by real life situations (where MB statistics dominates) and can run free, the more the BE statistics will appear.

What about the Fermi-Dirac (FD) statistics, can we also find traces of it in the human conceptual realm? We can observe that the interfaces with which the human concepts interact, i.e., the memory structures sensitive to their meanings, are certainly organized according to \emph{Pauli's exclusion principle}. Take the simple example of a computer, which will not allow one to make a copy of a file and name it in the same way, if memorized in the same folder. So, we can have identical copies of a same concept, but these identical copies must be in different states within the memory (in different folders in the computer). But we also know that entities formed by ordinary (baryonic) matter, which according to the conceptuality interpretation are the cognitive/memory-like entities interfacing with the bosonic messengers, are made of elementary fermions. So, one would expect to be also able to identify the equivalent of these fermionic elementary entities within our human conceptual realm. Consider for instance the well-known distinction between \emph{count nouns} and \emph{mass nouns} (also called \emph{non-count nouns}). The count nouns are those that can be combined with a numeral (and therefore also accept the plural form). They give rise to combinations like \emph{Ten Animals}, which as we discussed expresses a reality of ten identical entities all in the same state, typical of bosonic matter in so-called Bose-Einstein condensates. On the other hand, the mass nouns are those that have the property of not (meaningfully) combining with a numeral, without additional specifications. This means that we cannot have many identical non-count noun-concepts all in the same state. Take the example of the concept \emph{Courage}, whose associated word has no plural form. The combination \emph{Two courage} is clearly meaningless, which means that within the human language \emph{Courage} is not a boson-like conceptual entity, as we cannot put a given number of them all in the same state. We are however allowed to write combinations like \emph{Courage, courage, courage}, as we do when we repeat a word as a rhetorical device.\footnote{This is called an epizeuxis (or palilogia), and is typically used for vehemence, or emphasis.} But then there will be an order, which means that the \emph{Courage} conceptual entities in the combination will be in different states, which is the reason why the more they are in the combination, the greater will be the space required to write them on a page in the form of words. 

To push this parallel a bit further, consider also the combination \emph{Man of courage}. Even if it contains the non-countable concept \emph{Courage}, it can now be meaningfully combined with a numeral, for instance in: \emph{Ten men of courage}. This means that by combining a non-countable concept with other concepts, an emergent boson-like behavior can be obtained. This is similar to the well-known fact that fermions, when they aggregate, can behave as bosons, like in the typical example of the $\alpha$-particles (Helium nuclei). Note that fermions can become bosons only when they are bond by some kind of interaction, which as we know is in turn mediated by bosons. This means that, strictly speaking, fermions alone cannot form a boson: we cannot construct bosons without bosons. In the above combination, \emph{Man} is a boson-like (countable) concept, whereas \emph{Of} and \emph{Courage} are not. So, we could say here that the two fermion-like concepts \emph{Of} and \emph{Courage} interact through the boson-like concept \emph{Man}, producing the combination \emph{Man of courage}, whose behavior is boson-like. All this is of course for the time being purely heuristic, as we cannot expect to find within the human language conceptual realm the same level of organization of the microphysical realm (nor by the way we should expect that the former will necessarily evolve, in a far distant future, towards a same type of organization of the latter). In that respect, consider that the fermionic/bosonic duality of the micro-entities is intimately related to the rotational properties of the fractionary/integer spins they carry, according to the well-known \emph{spin-statistics theorem}. However, quoting from \cite{Aerts2009a}: ``[$\cdots$] although we can express the requirement of identity in general terms, the situation of human concepts and their interface of memory structures has not evolved sufficiently to contain a structure where rotational invariance may be expressed in general terms. This is also the reason that no equivalent of spin exists on this level.'' This does not mean that internal structures playing the same role in human concepts as spin and rotational invariance could not be identified, but this is a matter of future investigations.
 
To conclude this section about indistinguishability, consider also the concept \emph{Animals}, i.e., \emph{Animal} in the plural form, but not in a specific combination with a numeral. It clearly describes an ensemble of \emph{Animal} conceptual entities all exactly in the same state, but whose number is perfectly undetermined. If we write \emph{Animals} in an unpacked form, it can be understood as the infinite combination: \emph{One animal or two animals or three animals or four animals, etc.}, which in the Hilbert space mathematical language one would write as a coherent superposition of the states \emph{One animal}, \emph{Two animals}, \emph{Three animals}, etc., corresponding to the different possible numbers of \emph{Animal} conceptual entities in their ground state. If you think of the harmonic oscillator, this would be like a state $|\phi\rangle$ which is an infinite superposition of number-operator ($N=a^\dagger a$) eigenstates: $|\phi\rangle=\sum_{n=1}^\infty e^{i n\phi}|n\rangle$, i.e., a state where, according to the \emph{number-phase uncertainty relation}, the indetermination on the number of entities would be maximal, whereas the indetermination on their phases would be minimal, so much so that a description as a classical wave phenomenon would be possible. This is not the case for fermionic (non-countable) entities, for which, as is well-known, a classical undulatory approximation has no validity \cite{LeblondBalibar1984}.

\section{Measurement problem\label{measurement}}

In the previous sections, we have considered different quantum phenomena and explained how they can be understood in the light of the conceptuality interpretation. By doing so, we have described the measuring apparatuses as memory structures sensitive to the meaning carried by the measured quantum conceptual entities, so that measurements would be like interrogative contexts during which a conceptual entity, usually prepared in an abstract (superposition) state, is forced to acquire a more specific state, corresponding to one of the possible answers that the experimental setting permit to be selected (similarly to when we have to fill a multiple choice form having predetermined answers to select). Of course, the fact that a measurement is like an interrogative process is a metaphor which can be used independently of the conceptuality interpretation. Indeed, a scientist, by means of a measurement, certainly interrogates the system subjected to it, and the outcome is the answer it receives. But this is a description only at the human cognitive level, which is necessarily always present in a scientific experiment, as is clear that science is a human activity. The conceptuality interpretation, however, adds a new cognitive layer: that of the meaning driven interaction between the measured entity and the measuring apparatus. So, the following question arises: Can human decision processes shed a light into what happens behind the scenes of a quantum measurement process and provide an additional argument in favor of the conceptuality interpretation? 

To answer the above question, we have first to identify what are the important elements characterizing an interrogative process, when a cognitive entity is asked (or forced) to provide an answer when confronted with a given situation (that we can represent as a conceptual entity in a given state), selecting it from a number of predetermined possibilities. For this, we can ask what we intuitively feel when we are confronted with interrogative/decisional contexts of this kind. What we certainly can all recognize is that there will be a first phase during which we mentally immerse the situation in question into the context of the set of possible answers we have been given. If the situation is initially described by a state $D=|\psi\rangle\langle\psi|$ (which we write here as a projection operator), we can understand this first phase as a deterministic preparation process during which we bring the meaning of the situation as close as possible to the meaning of the different possible answers, which of course can also be described as final states of the conceptual entity which is the object of the interrogation. Assuming that there are $N$ possible answers, let us call $D_i=|\psi_i\rangle\langle\psi_i|$, $i=1,\dots,N$, these possible outcome states. So, there is a first immersive process during which the initial state $D$ will transition to a new state $D_e$, expressing this more specific meaning-connection with the different possible outcomes/answers $D_i$. Since only one of them can be selected (they are mutually excluding answers), this $D\to D_e$ immersive process creates a temporary state of unstable equilibrium (whence the index ``e'') between the competing tensions resulting from the different meaning-connections between $D_e$ and the $D_i$. Therefore, a second (usually indeterministic) phase will occur, which we can also subjectively perceive. It is the phase during which the mental tensional-equilibrium that was built is all of a sudden disturbed, in a way that cannot usually be predicted in advance, with the disturbance causing an irreversible process during which the conceptual state $D_e$ is drawn towards one of the possible answers $D_i$. This is really like a (weighted) symmetry breaking process, reducing the previously competing tensions and so allowing the cognitive entity to actualize an answer. 

Note that the above two-phase cognitive process is a general description that can account also for situations where the answer is known in advance. In this case, the tensional equilibrium that is built will be a trivial one, in the sense that the meaning-connection with one of the outcomes will always prevail and produce the predetermined outcome without fail. But of course, in the general situations the interrogated person will not have yet formed a strong opinion regarding which answer is to be selected, so that all answers can truly play a competing role in the creation of the tensional equilibrium, and will therefore have a non-zero probability to be selected. It is important to say that what we are describing here is really a model of the mind's processes and not a model of the brain's processes, and that of course mind and brain processes need not to be the same.\footnote{For example, the modeling of the activity of Broca's area is very different from the modeling of how human language is used, although of course there will be correlates.} But since the conceptuality interpretation assumes that the measuring apparatuses behave like cognitive entities, and the measuring apparatuses are precisely what physicists use to actualize an outcome, the following question arises: Can we also describe a quantum measurement process as a two-phase cognitive-like process where the initial state of the measured entity is first brought into a state of tensional equilibrium, which is subsequently broken in a way that the process exactly obeys the predictions of the Born rule? The answer is affirmative and the description in question is contained in the so-called \emph{general tension-reduction} (GTR) model \cite{AertsSassoli2015a,AertsSassoli2015b,AertsSassoli2016c}, which in the special case where the state space is Hilbertian and the measurements are uniform reduces to the \emph{extended Bloch representation} (EBR) that we mentioned already in Sec.~\ref{entanglement} \cite{AertsSassoli2014,AertsSassoli2016b,AertsSassoli2017} .

More precisely, there is a way to reformulate the standard quantum formalism by using a generalization and extension of the historical three-dimensional Bloch sphere model, which contains an exact description of the above two-stage process. In other words, the quantum formalism naturally generalizes and extends into a representation which is compatible with a general description of a measurement as a cognitive-like interrogative process. When we say that it generalizes the Bloch sphere model, it is because it applies to quantum systems of any dimension $N$, in fact also of infinite dimension \cite{AertsSassoli2017c}, and when we say it extends the Bloch sphere model, it is because it allows for a description, in the same representation, of the (hidden) measurement-interactions that are responsible for the breaking of the tensional equilibrium. It is of course not the purpose of the present work to enter into all the mathematical details of the GTR-model or the EBR of quantum mechanics. But let us provide some additional information about how the latter works. By introducing a representation for the generators of $SU(N)$, the special unitary group of degree $N$, it becomes possible to associate real $(N^2-1)$-dimensional unit vectors to the initial state $D$ and the final states $D_i$, which we will denote ${\bf r}$ and ${\bf r}_i$, $i=1,\dots,N$, respectively. These are vectors living at the surface of a convex region of states that is inscribed in a $(N^2-1)$-dimensional unit sphere $B_1(\real^{N^2-1})$, which coincides with the latter only in the two-outcome ($N=2$) case [thanks to the isomorphism between $SU(2)$ and $SO(3)]$. Now, one can show that the $N$ vectors ${\bf r}_i$ are the vertex vectors of a $(N-1)$-dimensional simplex $\triangle_{N-1}$, inscribed both in the convex region of states and in $B_1(\real^{N^2-1})$.

\begin{figure}
\begin{center}
\includegraphics[width=14cm]{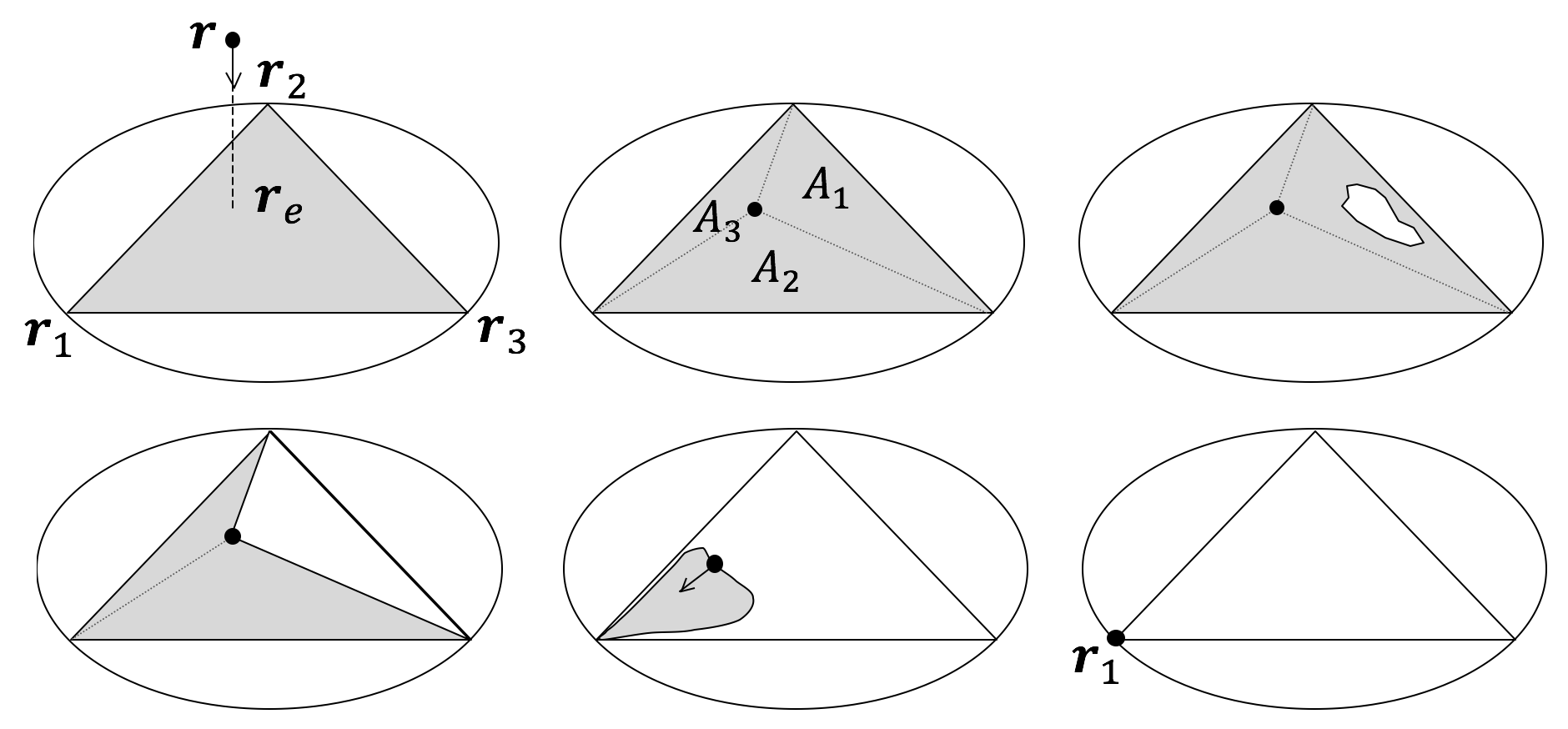}
\end{center}
\vspace{-15pt}
\caption{The unfolding of a measurement as a tension-reduction process, here with three possible (non-degenerate) outcomes: ${\bf r}_1$, ${\bf r}_2$ and ${\bf r}_3$. The abstract point particle representative of the initial state is positioned in ${\bf{r}}$, at the surface of the eight-dimensional sphere $B_1(\real^{8})$ (which of course cannot be drawn). It then orthogonally ``falls'' onto the triangular elastic substance $\triangle_{2}$ (an equilateral triangle) generated by the three outcomes, reaching the point ${\bf r}_e$ and so defining three convex sub-regions: $A_1$, $A_2$ and $A_3$. The substance of $\triangle_{2}$ then starts disintegrating at some unpredictable point, here inside $A_1$, so that $A_1$ fully disintegrates and detaches from its two anchor points, thus drawing the point particle to its final location, here ${\bf r}_1$.} 
\label{Figure9}
\end{figure}

The first phase of the measurement then corresponds to an immersion of the state vector ${\bf r}$ inside the sphere, along a path that is orthogonal to $\triangle_{N-1}$, reaching in this way an equilibrium point ${\bf r}_e\in \triangle_{N-1}$. This is the mathematical counterpart of the stage we previously described as the cognitive activity bringing the conceptual entity in full contact with the ``potentiality region'' generated by the $N$ mutually excluding answers. From a mathematical viewpoint, this causes the initial projection operator $D$, associated with ${\bf r}$, to gradually decohere and transform into a fully reduced density operator $D_e=\sum_{i=1}^N P_{\rm B}(\psi\to\psi_i) D_i$, associated with the (non-unit) on-simplex vector ${\bf r}_e$, where the positive numbers $P_{\rm B}(\psi\to\psi_i)=|\langle\psi_i|\psi\rangle |^2$ are the Born probabilities. And this means that in the EBR also the density operators play a role as representative of \emph{genuine states} (as we mentioned already in Sec.~\ref{entanglement}, in relation to the description of entangled sub-systems), describing the (non-unitary) evolution of the entity during the measurement itself. At this point, we can consider the ``tension lines'' going from the on-simplex state ${\bf r}_e$ to the $N$ outcome states ${\bf r}_i$, partitioning $\triangle_{N-1}$ into $N$ convex subregions $A_i$, formalizing the unstable tensional equilibrium we previously described. We can imagine these $N$ regions to be filled with an abstract elastic and disintegrable substance, so that when one of the regions -- say region $A_i$ -- starts disintegrating in a given internal point (this is the disturbance we previously described, due to the unavoidable fluctuations that are present in a measurement context), the disintegrative process will propagate within it, so that its $N-1$ anchor points will detach, with the consequence that the equilibrium state ${\bf r}_e$ (we can imagine it as an abstract point particle attached to the elastic substance) will be brought towards the remaning vertex vector, here ${\bf r}_i$, thus producing the measurement outcome (see Figure~\ref{Figure9}, for the $N=3$ case).
It then follows from the geometric properties of the structures involved that if we calculate the probability that the disintegration point happens in sub-region $A_i$, which is simply given by the ratio ${\mu(A_i)\over\mu(\triangle_{N-1})}$ between the $(N-1)$-dimensional volume (or Lebesgue measure) of sub-region $A_i$ and that of the full simplex $\triangle_{N-1}$, that such ratio is exactly given by the probabilities $P_{\rm B}(\psi\to\psi_i)=|\langle\psi_i|\psi\rangle |^2$, i.e., by the quantum mechanical Born rule \cite{AertsSassoli2014,AertsSassoli2015a}.

Let us mention that the process we just described, and its mathematical modeling, also generalizes to the situation of degenerate measurements (see the above references), when the \emph{tension-reduction process} does not result in a full resolution of the conflict between all the competing answers, so that the state is brought into a state of sub-equilibrium, between a reduced set of possibilities, described by a lower-dimensional sub-simplex of $\triangle_{N-1}$. To conclude this section about quantum measurements, let us also consider what could be a possible objection regarding our parallel between measurements in physics laboratories and cognitive processes where a mind-like entity selects one among a set of possible answers, according to the information stored in its memory. As we know, when we answer a question, the \emph{way} we do so can vary every time, depending on the mental state we are at that moment. Also, the way of choosing an answer of a person will generally differ from the way of choosing of another person. On the other hand, a measuring apparatus always 
chooses in the same way, which is the way described by the Born rule. In other words, each person should be associated with quantum-like probabilities which 
will generally differ from those predicted by the Born rule. This is of course correct, and as we mentioned already, we should not think of human culture, and the cognitive processes associated with it, as a reality domain that would have reached the level of symmetry and of organization of the micro-physical domain. 

But to be truthful, we don't really know if the measuring apparatuses always choose according to the Born rule. All we know is that the Born rule emerges from the statistics constructed from numerous outcomes. We thus cannot exclude that at each run $j$ of a measurement the apparatus would select an outcome according to probabilities $P^{(j)}(\psi\to\psi_i)$ which would generally differ from the Born probabilities $P_{\rm B}(\psi\to\psi_i)$. This would mean that an apparatus not only actualizes an outcome from a set of potential ones, but also, at a deeper level, actualizes a way of choosing an outcome from a (typically infinite) set of potential ways of choosing. Of course, for this to be consistent with the results we usually observe in the labs, the average $\langle P(\psi\to\psi_i) \rangle = {1\over N}\sum_{j=1}^n P^{(j)}(\psi\to\psi_i)$ should tend to the Born probability $P_{\rm B}(\psi\to\psi_i)$, as $n\to\infty$, for all $i=1,\dots,N$. This kind of average, called a \emph{universal average}, can be studied in the GTR-model by considering abstract non-uniform substances disintegrating in all possible ways, thus giving rise to all possible sets of probabilities for the different outcomes. These different non-uniform substances would describe the different ``mental states'' of the apparatus at each run of the measurement, and the remarkable result is that one can show that a universal measurement (when the state space is Hilbertian), exactly corresponds to a uniform measurement described by the quantum mechanical Born rule \cite{AertsSassoli2014,AertsSassoli2015b,AertsSassoli2017}.\footnote{See also \cite{AertsSassoli2014b}, for a discussion of the notion of universal average in relation to Bertrand's paradox.}

\section{Relativity}
\label{relativity}

To continue our exploration of the fertility of the conceptuality interpretation in providing new ways of explaining fundamental physical phenomena, we will now address \emph{relativity theory}. Indeed, not only the quantum phenomena, but also the relativistic ones, do challenge our classical prejudices and, as we are now going to explain, also for them the conceptuality interpretation can help us shed some light on their possible origin. In doing so, we will limit ourselves to consider the phenomenon of \emph{time dilation}. Also, we will limit our discussion to non-quantum relativistic entities (classical bodies) and will just provide at the end some clues about how to extend the reasoning to the quantum domain as well. But to begin with, let us observe that although the term ``relativity'' has been historically attached to Einstein, it refers in fact to a principle (the \emph{relativity principle}) that is much more ancient, as it was already described by Galileo Galilei in his famous example of the ship advancing at uniform speed, with people locked in the cabin beneath the deck not able to determine whether the ship was moving or just standing still \cite{Galileo1632}. In fact, one also finds descriptions of this principle as early as the first century B.C., i.e., 1700 years before Galileo, in China, in \emph{The Apocryphal Treatise on the Shang Shu Section of the Historical Classic: Investigation of the Mysterious Brightnesses} (\emph{Shang Shu Wei Kao Ling Yao}), where one can read: ``Although people don’t know it, the earth is constantly moving, just as someone sitting in a large boat with the cabin window closed is unaware that the boat is moving.''

A possible synthetic statement of the relativistic principle is as follows: ``Equivalent viewpoints exist on the physical world.'' When the principle is formalized by using the notion of \emph{reference frame}, it then becomes \cite{Levy-Leblond1977}: ``Equivalent frames of reference (space-time coordinate systems) exist for the physical laws, i.e., such that the physical laws have exactly the same form in all of them.'' This does not mean, however, that the different physical quantities will have the same values in the different equivalent reference frames: it means that they will obey exactly the same relations, so that phenomena will be perceived in the same way when experienced from these different but equivalent reference frames.\footnote{Of course, not all reference frames are equivalent. For example, when we are on a carousel rotating at a given speed, we will experience phenomena that would be absent if the carousel would be at rest, like the \emph{centrifugal pseudo forces}. The interesting content of the principle of relativity is therefore that among the countless possible reference frames, some non-trivial ones exist that are perfectly equivalent.} The simplest examples of equivalent frames of reference are those that are translated or rotated with respect to each other, but Galileo, and before him the ancient Chinese sages, identified a more interesting non-trivial class of equivalent reference frames: those moving with respect to each other at constant speed, called \emph{inertial frames}. The remarkable consequence of inertial frames that are equivalent frames is that an object moving at constant speed, from the viewpoint of the laws of physics, must be described in exactly the same way as an object at rest, i.e., as an entity on which the resultant force acting on it is zero. The \emph{first law of Newton}, or \emph{principle of inertia} then immediately follows: an object in motion at constant speed, like an object at rest, will forever remain in such state of motion, if not acted upon by some additional force. 

A much more remarkable consequence follows from the observation that there are wave phenomena (like the electromagnetic ones) that appear to propagate through the very ``substance of space,'' once called the ether. Indeed, if this would be the case, i.e., if space would be substantial and waves could propagate through its medium, then some physical effects (like interference effects) should manifest differently in different inertial frames, thus contradicting the very relativistic principle. But if the principle is true, as it appears to be, these differences should not be observed, and in fact have not so far been observed, for instance in the historical Michelson-Morley experiment and in those that followed, which showed instead that the speed of propagation in space of the electromagnetic fields is always the same, for all inertial frames and in all directions. This means that space, understood as an encompassing substantial theater for reality, becomes a problematic notion and that what we call space is essentially a relational construct, so that each physical entity, with its unique perspective, would actually inhabit `a different space'. And this means that, as it will become clearer in the following, we do not see objects moving in space because they would actually move in an objective spatial theater, but because we confer them a movement in order to keep them inside our personal spatial representation. Now, as is well-known, when the relativistic principle is applied in conjunction with some very general and natural hypothesis about space and time, the Lorentz transformations are obtained as the only possible transformations connecting the different equivalent inertial frames \cite{Levy-Leblond1976}. Remarkably, these transformations do not affect only the spatial coordinates, but also the temporal ones, and the consequence is that objects, when they move with respect to a given reference frame, they are shorter in comparison to when they are at rest (\emph{length contraction}), and also, objects called clocks, when they move also run more slowly in comparison to clocks that are at rest (\emph{time dilation}). This means that what relativity is telling us is that the spatial constructs associated with the different physical entities cannot be just spatial, but have to be genuinely spatiotemporal. 

To highlight this fact, consider the following thought experiment (see \cite{Aerts1999} for a more extensive discussion). Imagine that you are at the \emph{Vrije Universiteit Brussel} (VUB), in Belgium (usually referred to as the \emph{Free University of Brussels}, in English-speaking contexts), and that it is September 29, 2017, say 3 pm.\footnote{On September 29--30, 2017, the \emph{Centre Leo Apostel for Interdisciplinary Studies} (CLEA), Belgium, has organized the international symposium ``Worlds of Entanglement,'' during which one of the authors presented the guidelines of the conceptuality interpretation to an heterogeneous audience, formed not only by physicists, but also mathematicians, social scientists, biologists, artists, philosophers, economists, and others. The present article, is an extended version of the content of that presentation.} We can call this your personal present moment $t_0$. When you are at VUB, at time $t_0$, since you are having a direct experience with the university, you can affirm with certainty that VUB is real for you, i.e., that VUB is an existing element of your present personal material reality. But what about the reality of, say, the \emph{Universit\'a della Svizzera italiana} (USI), in Switzerland (usually referred to as the \emph{University of Lugano}, in English-speaking contexts)? Since at time $t_0$ you are at the VUB, and you are not having an experience with the USI, can you nevertheless affirm that the USI is also an element of your present personal reality, at time $t_0$? The answer is positive, and the reason for this is that, following EPR's reality criterion,\footnote{In a famous article written in 1935, Einstein, Podolsky and Rosen (EPR) recognized that our construction of reality is based on our predictions about it. The original wording of their criterion is \cite{Einstein1935}: ``If, without in any way disturbing a system, we can predict with certainty [...] the value of a physical quantity, then there exist an element of physical reality corresponding to this physical quantity.'' For a discussion of the criterion, see \cite{sdb2011} and the references cited therein.} we know that reality is a construction about the possible: if, in your past, you would have decided to travel to Lugano, Switzerland, then with certainty you would have had a direct experience with the USI at the present time $t_0$, and considering the certainty of such a prediction, you can say that also the USI is an element of your personal reality, at time $t_0$. Consider now the VUB at subsequent time $t_1 >t_0$, where $t_1$ is September 30, 2017, 3 pm, i.e., one day in your future with respect to your present time $t_0$. Is the VUB at time $t_1$ also an element of your reality? If we rely only on our parochial conception of space and time, we would respond negatively, but this would be a wrong answer considering what we know about the relativistic effects, and more specifically the effect of time dilation: the slowdown of the ticking rate of moving clocks, when compared to those that remain at rest.

\begin{figure}
\begin{center}
\includegraphics[width=10cm]{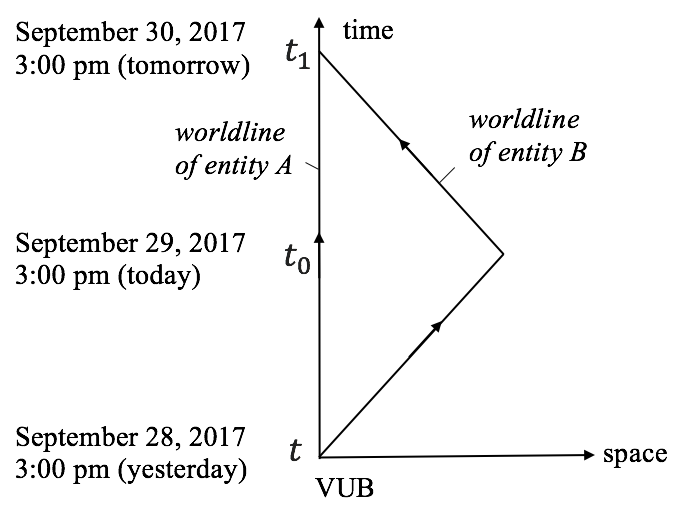}
\end{center}
\vspace{-15pt}
\caption{The two worldlines of the entities $A$ and $B$, in the spacetime construction associated with the former. Entity $A$ is spatially at rest, thus only moves along her time axis, whereas entity $B$ goes on a round-trip journey, allowing her to meet again with entity $A$, in her -- one day after -- future.} 
\label{Figure10}
\end{figure}

Indeed, if in your past, say on September 28, 2017, 3 pm, you would have used a space ship to travel at speed $v=\sqrt{3\over 4} \, c$ (where $c$ is the speed of light) to any destination, then back again along the same route, because of the relativistic time dilation effect you could have been back at VUB exactly when your smartphone would indicate September 29, 2017, 3 pm, whereas the smartphones of all other people at VUB would indicate September 30, 2017, 3 pm. So, if you take seriously EPR's reality criterion, you must conclude that the VUB, one day in its future, is also an element of your present personal reality. Now, since the present discussion is aimed at an interdisciplinary audience, we think it can be useful to also briefly explain how time dilation is calculated in relativity theory. So, there are two versions here of the same individual, one remaining at rest at VUB,\footnote{VUB being on the surface of planet Earth, strictly speaking it cannot be associated with an inertial frame, but for simplicity we will neglect the planet's non-uniform motion in our reasoning.} who we will call entity $A$, and the other performing the round-trip journey, who we will call entity $B$ (see Figure~\ref{Figure10}). If we denote $\textsc{t}_B$ the time-period of the clock carried by $B$ during her/his trip, as measured by $A$, using an identical clock remaining at VUB, the time period of which is $\tau_A$,\footnote{Note that we are using a different notation for the two time periods $\tau_A$ and $\textsc{t}_B$. This because the former is a so-called \emph{proper time}, i.e., a time measured by a clock which remains at rest with respect to $A$, whereas the latter is a \emph{coordinate time}, i.e., a time measured by a clock which is not at rest with respect to $B$.} s/he will observe a time dilation effect, i.e., that $\textsc{t}_B$ is greater than $\tau_A$. More specifically, if $v$ is the speed of $B$ (when moving away or approaching $A$), then we have $\textsc{t}_B=\gamma\, \tau_A$, where $\gamma=1/\sqrt{1-{v^2\over c^2}}$ is the so-called \emph{Lorentz gamma factor}, which is equal to $2$ for the above value of the speed $v$. Hence, we have that $\textsc{t}_B=2\tau_A$, i.e., that the clock traveling with $B$ appears to $A$ to run twice as slow than the clock that remained at VUB. 

Let us now assume that $A$ measures $n_A$ cycles of her/his clock for the entire duration of the trip of $B$.\footnote{For simplicity of the discussion, we will neglect that there are also accelerations experienced by $B$, at her/his departure, turnaround and arrival.} Since $\textsc{t}_B=\gamma\, \tau_A$, the number of cycles $n_B$ of the clock of $B$ will be obtained by solving the equation: $n_A\tau_A=n_B\textsc{t}_B=n_B\gamma\, \tau_A$, which gives $n_B={n_A\over \gamma}$, and for our value of the speed $v$ we have: $n_B={n_A\over 2}$. In other words, the traveling entity $B$ 
uses half the time-cycles of the non-traveling entity $A$. Now, to determine the time $t<t_0$ (where $t_0$ corresponds to September 29, 2017, 3 pm) at which $B$ would have needed to start her/his space travel at speed $v=\sqrt{3\over 4} \, c$, in order to be back at the same place, at VUB, at time $t_1$ (corresponding to September 30, 2017, 3 pm), with her/his clock indicating September 29, 2017, 3 pm, we can reason as follows. By definition, we have $n_A={t_1-t\over \tau_A}$, and let us also denote $n'_A$ the number of cycles corresponding to a one day (24 hours) period: $n'_A={t_1-t_0\over \tau_A}$. We want that $n_B=n_A-n'_A$, i.e., we want the clock of $B$ to use 24 hours less than the clock of $A$. Since $n_B={n_A\over \gamma}$, we obtain $n_A={\gamma\over \gamma -1}\, n'_A$, so that for $\gamma =2$ we have $n_A=2n'_A$. In other words, $B$ has to start her/his trip two days before September 30, 2017, 3 pm, that is, on September 28, 2017, 3 pm (see Figure~\ref{Figure10}).

Coming back to our discussion, being our personal present reality defined in a counterfactual way, via the EPR criterion, we have to conclude, as a consequence of the relativistic \emph{generalized parallax effects}, that our personal present also contains a part of our personal future. However, this not in the sense that all of our future would be given, as if the universe would be an unchanging block. Indeed, if it is true that in a given reference frame we can always attach time and space coordinates to the different events, this doesn’t mean that the processes of change that have created them are also happening in space and time. Indeed, these processes typically originate from a non-spatiotemporal realm, which remains hidden from our limited spatiotemporal perspective. So, if Galilean relativity has told us that physical entities are not inhabiting a substantive objective space, as each entity constructs a personal three-dimensional relational space, Einsteinian relativity has pushed such view a step further, telling us that entities are not only not inhabiting a substantive space, but also that they do not construct their time axis in the same way, i.e., that each entity constructs a personal four-dimensional spacetime. We thus see that, similarly to quantum mechanics, relativity also indicates the existence of an underlying non-spatial and non-temporal realm. And as we are now going to explain, the hypothesis that physical entities would have primarily a conceptual nature is not only able to offer an explanation for the strangeness of the quantum effects, but also for the relativistic ones, which are erroneously considered to be less strange than the former (if we try to understand them by maintaining a purely spatiotemporal perspective).

\section{Time dilation}
\label{dilation}

Let us consider again the previous example, assuming this time that $A$ and $B$ are not two different versions of the same person, who made a different choice in the past, but two different physical entities, so that we are now in the specific situation of Langevin's twin-paradox. Note that the reason why it was referred to as a paradox is the fact that one could argue that by considering the viewpoint of the reference frame associated with the space ship, it is the entity remaining on Earth that appears to have performed the return trip. This apparent symmetry between the two descriptions is however broken as soon as one observes that the two reference frames are non-equivalent, as is clear that the frame associated with entity $B$, using the space ship, is a non-inertial one. In other words, the symmetry is broken by observing that $B$ experiences accelerations that are not experienced by $A$ (neglecting those associated with the rotation of the planet). One should not conclude, however, that the observed time-dilation effect (or length contraction effect, from the viewpoint of the traveling entity) would be caused by these accelerations: it is in fact the geometric structure of the worldlines associated with the two entities that is responsible for the time dilation, which is truly defined by the Lorentz-invariant length corresponding to the so-called \emph{proper time} interval associated with them \cite{Aerts2017}.

The two entities $A$ and $B$ are here considered to be classical macroscopic bodies, i.e., ordinary objects. However, as we discussed in Sec.~\ref{objects}, in the conceptuality interpretation objects are idealizations of story-like conceptual entities that can be in different meaning-states. So, we want now to consider the two entities $A$ and $B$ not as objects moving in space but, primarily (and more fundamentally), as conceptual entities that can have meaning driven interactions. In relativity, one usually associate observers with entities in different states of motion, where the notion of \emph{observer} is typically understood as a shortcut for a reference frame plus an entity that, if it would be present in some specific location, would be able to perceive (detect, measure) phenomena relative to the viewpoint of that reference frame and specific location.\footnote{To quote a passage from \cite{Einstein1920} (emphasis is our): ``If the observer \emph{perceives} the two flashes of lightning at the same time, then they are simultaneous.''} We will also associate observers with the two entities $A$ and $B$, but we will consider them as mind-like entitities sensitive to the meaning carried by $A$ and $B$. Let us simply call them \emph{cognitive observers}, and denote them $C_A$ and $C_B$. These two observers are however not associated with spatiotemporal frames of reference. The only aspect distinguishing $C_A$ from $C_B$ is that the former is focused on the evolution of $A$, whereas the latter is focused on the evolution of $B$. 

To fix ideas, we can simply consider that the process of change of state of entity $A$ corresponds to the cognitive activity of entity $C_A$, reflecting on a given problem, so that the initial state of $A$ would correspond to the \emph{Hypothesis} initiating such reflection, and the final state of $A$ to the \emph{Conclusion} reached by $C_A$, after having followed a certain number of intermediary \emph{conceptual steps}. And same for the cognitive observer $C_B$, following the evolution of the conceptual entity $B$.\footnote{This means that we are here considering $A$ and $B$ to correspond to the conceptual entities \emph{Reasoning of} $C_A$ and \emph{Reasoning of} $C_B$, respectively.} Here we will assume that $C_A$ and $C_B$ are just witnessing the unfolding of the meanings carried by $A$ and $B$, as they evolve, i.e., that they are not themselves producing the observed changes of their states. Also, to place ourselves in the ``twin-paradox'' situation, we consider that $C_A$ and $C_B$ are both reflecting on the same problem, starting with the same \emph{Hypothesis} and subsequently jointly reaching the same \emph{Conclusion}. In other words, in the conceptual abstract realm that they both inhabit, they have a first meeting at the ``place'' of their commonly shared \emph{Hypothesis}, then a second encounter when they reach the same \emph{Conclusion}. The difference between $C_A$ and $C_B$, however, is that the cognitive path they follow to reach that same \emph{Conclusion}, starting from the same \emph{Hypothesis}, is not the same, in the sense that $C_A$, focused on the evolution of $A$, is assumed to use $n_A$ conceptual steps to do so, whereas $C_B$, focused on the evolution of $B$, is assumed to use a lesser number of steps $n_B<n_A$. Let us denote $A_i$, $i=0,1,\dots, n_A$, the different states through which $A$ passes to go from the \emph{Hypothesis} = $A_0$, to the \emph{Conclusion} = $A_{n_A}$, and let us denote $B_i$, $i=0,1,\dots, n_B$, the states $B$ transition through to also go from the \emph{Hypothesis} = $B_0$, to the \emph{Conclusion} = $B_{n_B}$.

Imagine then that the cognitive observer $C_A$, to keep track in an orderly manner of the conceptual path followed by entity $A$, decides to introduce an axis to parameterize each one of $A$'s conceptual steps. For this, it will ascribe a unit length $L_A$ to such axis, corresponding to a single conceptual step, and it will also assume that the speed at which each step is accomplished is the same for all steps and is equal to a given constant $c$, so that the duration of a single step is: $\tau_A={L_A\over c}$. When going from the \emph{Hypothesis} to the \emph{Conclusion}, the reasoning of $C_A$ will thus correspond to a movement of entity $A$, along such \emph{order parameter axis}, going from an initial point $D_0$ to a final point $D_{n_A}= D_0 + n_A L_A = c\, (t_0+n_A\tau_A)$, where we have defined the times $t_i ={1\over c}(D_0 + i L_A)$, $i=0, \dots, n_A$, where $t_0 ={D_0\over c}$ is the initial time and $t_{n_A}= {D_{n_A}\over c}$ the final time. Consider now the evolution of entity $B$, which we assumed can reach the same \emph{Conclusion} following a shorter conceptual path, only made of $n_B<n_A$ steps, and for simplicity we will consider here that $n_B={n_A\over 2}$. The cognitive observer $C_A$ can also decide to focus on the evolution of $B$, i.e., might also be willing to keep track of the cognitive path followed by entity $B$, in addition to that of $A$. Now, if $A$ and $B$ are entities of the same nature, it can be assumed that when they produce a cognitive step, they do so at the same speed $c$. But then, since the path followed by $B$ in the abstract conceptual realm is such that it can reach the same \emph{Conclusion} in half the steps used by $A$, the cognitive observer $C_A$ cannot represent such path on the same axis used to parametrize the path of $A$, as units on the latter were precisely chosen in a way that one needs twice the number of steps to reach the \emph{Conclusion}. 

\begin{figure}
\begin{center}
\includegraphics[width=14cm]{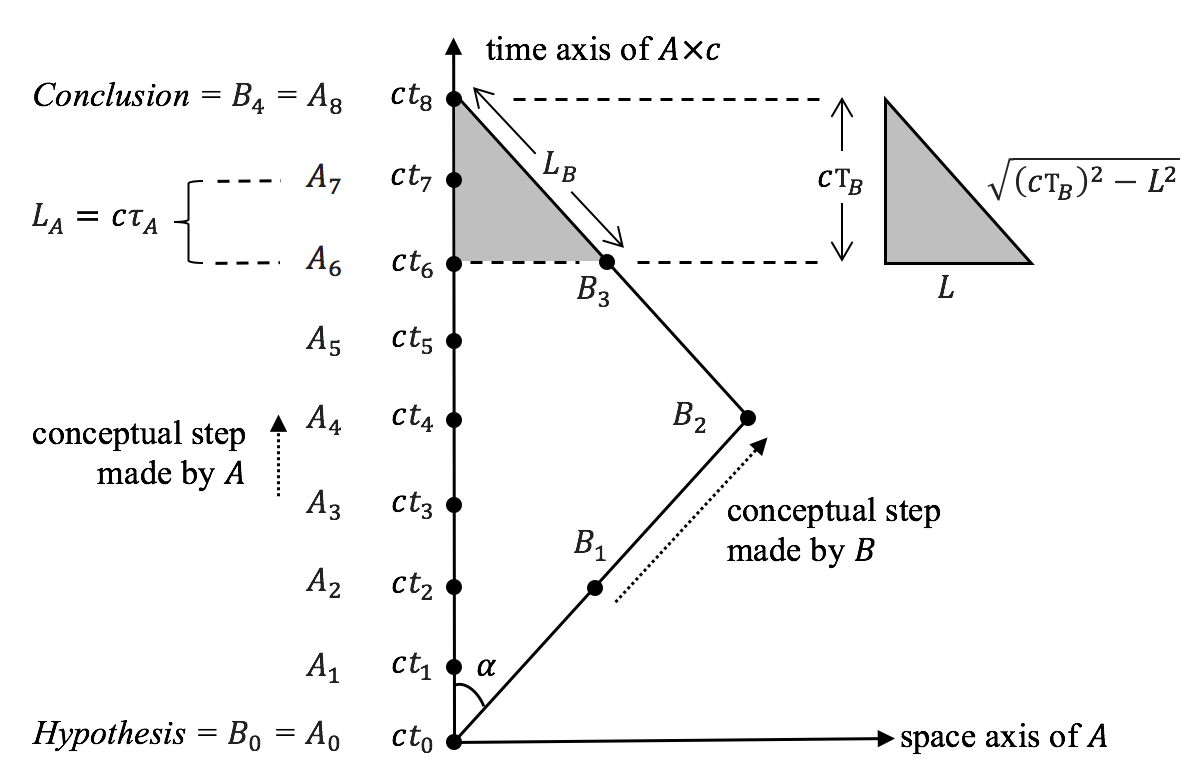}
\end{center}
\vspace{-15pt}
\caption{The coordination of the conceptual paths followed by the two entities $A$ and $B$, in the spacetime constructed by the cognitive observer $C_A$ (here in the situation $n_A=8$ and $n_B=4$). When measured along the time axis of $A$ (multiplied by the constant speed $c$) the length $L_B=c\, \textsc{t}_B$ of the conceptual steps of $B$ appear to be longer than the length $L_A=c\,\tau_A$ of those of $A$. However, when measured along the direction of its own movement in the $A$-spacetime, using the \emph{Minkowski} instead of the \emph{Euclidean} metric, one finds that the conceptual steps  of the two entities are exactly of the same length, in accordance with the fact that they both move at the same (absolute) constant speed $c$ in the underlying conceptual realm.} 
\label{Figure11}
\end{figure}

To consistently parametrize also the evolution of $B$, $C_A$ is thus forced to introduce an additional axis, and use the additional dimension generated by such axis to describe $B$ as moving on a round-trip path, now contained in a higher dimensional space generated by both the first parametric axis -- let us call it the \emph{time axis of $A$} -- and this second parametric axis -- let us call it the \emph{space axis of $A$}. So, the evolution of entity $B$ is described as a movement on a path leading away from such time axis and then coming back to it, to reach the \emph{Conclusion} meeting point, 
and this by doing exactly ${n_A\over 2}$ cognitive steps (see Figure~\ref{Figure11}). However, if we consider the construction of this parametric space from a purely \emph{Euclidean} perspective, we immediately see that things do not work. Indeed, if we calculate the length of the $B$-path using the \emph{Pythagorean theorem}, we will necessarily find a path that is longer than that walked by $A$. 

This would not be correct, as is clear that $B$ follows a shorter conceptual path, only using half of the conceptual steps used by $A$. Therefore, when measuring the length of $B$'s conceptual path, it should be shorter and not longer than that of $A$. For $C_A$ to fix this problem, the only way to go is to consider a \emph{pseudo-Euclidean} space, instead of an Euclidean one, and more precisely that specific pseudo-Euclidean space known as the \emph{Minkowski} space (or spacetime), where distances are not calculated using the \emph{Pythagorean theorem}, but a \emph{pseudo-Pythagorean theorem} attaching a negative sign to the squares of the components associated with the space axis, and a positive sign to the square of the components associated with the time axis. In this way, the length of the hypotenuse of a right triangle, whose catheti are associated with the time and space axes, respectively, will generally be less than the length of the time-cathetus. It becomes then possible for the length $L_B$ of a single conceptual step of B (see Figure~\ref{Figure11}) to be exactly equal to the length $L_A$ of a single conceptual step of $A$, i.e., to have the equality $L_A=L_B$, which is what $C_A$ wants to have, as the two entities $A$ and $B$ are assumed to change state at the same absolute speed $c$ (the speed of light in vacuum) in their common conceptual realm, so that the duration/length of their conceptual steps must be an \emph{invariant}, i.e., the same for all entities.

More precisely, if $L$ is the component of the length $L_B$ along the space axis of $A$, then according to the pseudo-Euclidean (Minkowski) metric we have: $L_B^2 =(c\,\textsc{t}_B)^2 -L^2$, so that the requirement that $L_A=L_B$, or equivalently $L_A^2=(c\,\tau_A)^2=L_B^2$, considering that $\tau_A={1\over \gamma} \textsc{t}_B$ and $c\,\tau_A={c\over\gamma}\,\textsc{t}_B= \sqrt{c^2-v^2}\,\textsc{t}_B$, gives: $(c^2-v^2)\,\textsc{t}^2_B =(c\,\textsc{t}_B)^2 -L^2$, that is, $L=v\textsc{t}_B$. In other words, by adopting a pseudo-Euclidean (Minkowski) metric, the cognitive observer $C_A$ is able to construct a spacetime theater in which it can keep track, in a consistent way, not only of the cognitive process associated with $A$, but also of that associated with $B$,\footnote{A single spatial axis is sufficient when considering only two entities. However, additional space axes are needed if further entities are considered; see \cite{Aerts2017}.} and to do so all it has to do is to attach an appropriate spatial velocity $v$ to characterize its state changes. In other words, the reason for the time-dilation generalized parallax effects becomes clear when the existence of an underlying conceptual realm is taken into consideration: since $C_A$ has to also parametrize the cognitive path of $B$, and cannot do it using the same time-axis, it has to consider a movement within a higher dimensional space, characterized by an angle $\alpha =\tan^{-1} {v\over c}$ with respect to the direction of the movement of $A$. This will inevitably introduce \emph{temporal effects of perspective}: $C_A$ will observe $B$ as if it was producing conceptual steps (or cycles) having an increased duration $\textsc{t}_B=\gamma \tau_A$. This means that $C_A$, focusing its attention on $A$, when it compares its cognitive activity with that of an observer $C_B$, focusing its attention on $B$, will have the impression that $C_B$ reasons more slowly than itself, but since it also reasons more efficaciously, as it uses a lesser number of conceptual steps, they are nevertheless able to meet at the common \emph{Conclusion} state. This is just how things appear to be at the level of the spacetime parametrical construction operated by $C_A$. At the more objective level of the non-spatiotemporal conceptual realm, $A$ and $B$ move at exactly the same speed $c$, which is the intrinsic speed at which they both perform their conceptual steps. 

Our description of time-dilation effects would of course require more explanations, and we refer the reader to \cite{Aerts2017}, where more details can be found. Our main point here was to highlight that relativity theory, similarly to quantum mechanics, indicates the existence of a non-spatiotemporal conceptual realm. As we mentioned already, our discussion indicates that its non-temporality is however not to be understood in the sense of an absence of processes of change. On the contrary, every conceptual (physical) entities would incessantly change state, i.e., produce new conceptual steps, by all ``surfing'' over the conceptual realm at the light speed $c$. Therefore, at a more fundamental level, movement would be incessant, and in a sense absolute. This is possible because it is not a movement in space and time, as space and time would only emerge when a cognitive observer decides to coordinate the evolution of a given conceptual entity with the evolution of other conceptual entities, introducing for this a specific Cartesian coordinate system. In such system, the time axis orders the conceptual changes of the entity the cognitive observer decides to primarily bring its focus to, whereas the spatial axes order the evolution of the other conceptual entities, relative to such proper time-axis, by representing them as movements in space. Such spatiotemporal construction, to be consistent, requires the metric to be Minkowskian, which of course remains counterintuitive to us humans, as we evolved on this planet by mostly interacting with entities moving extremely slowly in space with respect to one another, i.e., that are almost at rest with respect to one another, so that the relativistic parallax effects, being negligible, were not integrated in our mental representation of the world. 

Of course, this spatiotemporal representation only works for conceptual entities having reached the status of objects, the so-called classical macroscopic bodies. When micro-physical entities are considered, the time-space duality must be replaced by a more general duality between time and the set of outcome-states associated with the different possible measurements. This is for instance the situation where the cognitive observer $C_A$ would not merely witness the surfing of entity $A$ over the more fundamental conceptual realm, but in fact would also affect its surfing through its observation, thus also introducing in its evolution the additional ingredient of indeterminism. Note that the possibility for $C_A$ to also act as a quantum measurement context for entity $A$ is not incompatible with the special situation of a deterministic evolution. Indeed, any deterministic change of state can in principle be conceived as being the result of a measurement having just a single possible outcome.\footnote{Hence, two-outcome measurement processes would not constitute the simplest imaginable measurement situation.} This means that deterministic evolutions can in principle be described as recursive applications of multiple one-outcome measurement processes. Some of these processes will be governed by classical contexts, and the corresponding deterministic evolution can be described as an `evolution in space', others will be governed by genuine quantum contexts, and the corresponding deterministic evolution cannot be described as happening in space, but in a more abstract (conceptual) non-spatial (and non-temporal) realm. 

We already mentioned in Secs.~\ref{entanglement} and \ref{measurement} the extended Bloch representation (EBR) of quantum mechanics \cite{AertsSassoli2014,AertsSassoli2016b,AertsSassoli2017}, which can be used to construct a quantum theater in which all the measurement processes associated with a quantum entity, and its states, can be jointly represented. For measurement contexts admitting sets of up to $N$ possible outcome-states, the number of required dimensions for the associated Blochean quantum theater is equal to $N^2-1$, which is the number of generators of the $SU(N)$ group of transformations. Roughly speaking, these transformations can be interpreted as ``generalized rotations,'' and this means that to enter such Blochean theater one has in a sense to ``rotate away'' the intrinsic complexity of a quantum entity, by means of these generators. A human conceptual analogy here would be that of considering that to enter a given space of discourse, like that of a political agenda, certain concepts first need to receive a ``twist.'' Our spatiotemporal theater, considered as a specific space of discourse, would require in the same way specific ``twists'' to be applied, for the different quantum conceptual entities to enter and be representable in it.

We conclude our discussion about relativistic effects with a brief remark about gravitation. As is well known, we are still lacking a satisfactory quantum gravity theory, and this because the fundamental forces in the \emph{Standard Model} of particle physics are modeled as (quantized) fields in a fixed spatiotemporal background, whereas the gravitational forces precisely affect that background, making it a dynamical one. Different from the Standard Model and similar approaches, attaching a fundamental role to the spatiotemporal canvas, the conceptuality interpretation posits that reality is not contained in spacetime, the latter being just a relational construction emerging each time a very specific interface is considered: that between the macroscopic pieces of matter and the force fields acting on them, i.e., between the fermionic constructions and their bosonic way of exchanging meaning. It is in this interface that the illusion was formed of a spatiotemporal theater in which our physical reality would be fully contained; an illusion which was then consolidated through the very scientific experimental method, somehow forcing us to only approach our physical reality through such interface, as is clear that physicists, in their laboratories, always collect data from experiments involving apparatuses formed by macro-pieces of matter. If space and time (we should better say spaces and times) are by-products of this very specific interface, we can more easily understand the reason of the difficulties that were encountered in the attempts to construct a consistent quantum gravity theory. The conceptuality interpretation, by pointing to the existence of a more fundamental and abstract realm, in which the physical conceptual entities evolve, the fundamental forces, gravity included, are then allowed to be understood as expressions of the different ways conceptual entities can exchange meaning, and because of that be brought together, or apart.

\section{Conclusion}
\label{conclusion}

It is time to move towards the conclusion of our \emph{tour d'horizon} of the conceptuality interpretation and its explicative power. In this last section, we will just evocate some possible directions for subsequent investigations, and in this regard we also refer the interested readers to \cite{Aerts2009a,Aerts2010a,Aerts2010b,Aerts2013,Aerts2014}.

Concerning the so far failed tentative to unify gravitational and quantum elements of reality within a unique consistent theoretical construction, which we mentioned in the previous section, let us observe that the conceptuality interpretation brings another interesting line of reflection: it is also a possibility that a single `quantum plus gravitational' description might not be feasible, in the sense that `quantum' and `gravity' could very well be incompatible descriptions, in the same way that position and momentum measurements are incompatible experimental contexts. Indeed, a conceptual reality is also a contextual reality, i.e., a reality where certain meanings would be actualized and actualizable only in certain contexts, and not in others. In that respect, classical physics can also be understood as a description emerging from a very specific context, produced by us humans mostly manifesting and interacting with physical entities through our macroscopic dense bodies. Standard quantum mechanics, and more precisely its formalization through the Hilbertian formalism, can be considered as another context associated with different operationally posed questions, whose answers cannot be all organized in the `space of relations' that resulted from the previous classical construction, forming a sort of closed representational environment (somehow in the spirit of Heisenberg’s notion of \emph{closed theories} \cite{Bokulich2008}). But the quantum representation, which also has its structural shortcomings, might as well form another closed environment, considering for instance its inability to describe entities that can remain separated in experimental terms \cite{Aerts1984,Aerts2014,AertsSassoli2017e}. In other words, it is also possible that a single encompassing representation could not be obtained, precisely because it would correspond to the unrealistic desiderata of simultaneously actualizing properties/meanings that are in ultimate analysis associated with incompatible contexts.\footnote{This is a view that subtends a notion of realism that was recently introduced and called \emph{multiplex realism} \cite{AertsSassoli2017d,AertsSassoli2017e}.}

In addition to that, the conceptuality interpretation, with its hypothesis that physical entities are fundamentally conceptual, also fosters a \emph{pancognitivist} view (as was mentioned in the Introduction), where every element of reality would in fact participate in cognition, with human cognition being just a special case of it, appearing at a very particular organizational level. This has clearly deep consequences on our view about \emph{evolution} in general, as the advent of the biological species on our planet, including the human one, would only be part of a much wider and fundamental process of change resulting from the interaction of conceptual entities with the countless cognitive structures that are sensitive to their meaning, and this since the dawn of the formation of our universe and at different levels of the same. If this is correct, the default picture we should adopt in the description of our evolving physical reality is that of a huge and multilayered \emph{cultural evolution} \cite{AertsSassoli2017b}. So, in the same way we humans use concepts and their combinations to communicate and evolve our cultures, the same might have occurred, and would still be occurring, in the micro-realm, and this automatically provides a compelling explanation for so-called \emph{dark matter}, which can then be understood as that part of matter which, as an interface, has not co-evolved together with the bosonic ``messenger'' entities. Think of the abundance in our human environment of those structures that cannot exchange human meaning, i.e., the ordinary pieces of matter as opposed to the cultural artifacts, the former being much more abundant than the latter. The same could be for dark matter, as opposed to ordinary matter, which not only does not interact with the bosonic micro-carriers of meaning, but also appears to be indeed much more abundant. On the other hand, gravity, by working at a very different scale than all the other forces, would possibly describe a more ancient way of exchanging meaning and creating concentrations of it; a way which has remained in common with both ordinary and dark matter. 

This special role played by gravitation can also be seen in the diversity of the mass values of the different micro-physical entities, which are not just multiples of some fundamental unit, as it is the case for instance for the electric charge. This seems to suggest that mass is not so much connected to the notion of \emph{identity} of a given conceptual entity, but instead to the different possible ways a given identity is able to manifest. Think of the puzzling existence of the three different \emph{generations of elementary micro-entities}. Entities that are members of these different generations interact exactly in the same way, but differ in their masses. To give an example, there are three different electronic entities: the ordinary \emph{electron} of the first generation, having a mass of 0.511 MeV/${\rm c}^{2}$, the \emph{muonic} electron of the second generation, having a larger mass of 106 MeV/${\rm c}^{2}$, and finally the \emph{tauonic} electron of the third generation, with an even larger mass of 1777 MeV/${\rm c}^{2}$ (almost twice the mass of a proton). The conceptuality interpretation offers the following possible element of explanation for these different generations of micro-entities: they would simply correspond to different energetic realizations of a same conceptual entity, in the same way as in our human culture a concept can manifest as, say, a spoken sound-energetic form, an electromagnetic and/or electronic form, in a carved into stone form, etc., and all these different forms, although they have different mass-energies, they nevertheless always convey the same meaning, i.e., they interact in a meaning-driven environment in exactly (or almost exactly) the same way. 

Let us for a moment also mention the issue of the observed intrinsic expansion of the universe, according to current Big Bang theories. The recurring question is: ``In what the universe is expanding into?'' And the recurring answer is: ``This is a nonsensical question, as the universe contains everything and there is nothing into which it could be expanding, so, it is just expanding!'' Of course, this kind of answer is perceived as highly unsatisfactory to the layman, and rightly so, as we think it should be unsatisfactory to the professional physicist as well. As we discussed at length in this article \cite{Aerts1999}: ``Reality is not contained within space. Space is a momentaneous crystallization of a theatre for reality where the motions and interactions of the macroscopic material and energetic entities take place. But other entities -- like quantum entities for example -- `take place' outside space, or -- and this would be another way of saying the same thing -- within a space that is not the three-dimensional Euclidean space.'' The conceptuality interpretation allows one to push even further this statement, by observing that reality's non-spatiality results from it being of a fundamental conceptual nature, implying  the existence of a multilayered structure resulting from the interplay between states having different degrees of abstractness and concreteness. 

In other words, the expansion of our universe would simply be the result of a cosmic-cultural evolution constantly creating new stories (through a mechanism of conceptual combination), which emerge from a substrate of more abstract  entities (i.e., conceptual entities in more abstract states) that can combine together to form more complex states. To draw a parallel with the Web, think of the constant creation of new webpages, arising from the activity of all cognitive entities participating in the associated human meaning-driven interactions. In that respect, it is interesting to observe that the expansion of the Web, since the first web-site was published back in 1991, has been an accelerated one, so one can also think of the observed increasing rate of expansion of our universe to be the result, \emph{mutatis mutandis}, of a cultural accelerated growth mechanism. Let us also mention that a conceptual reality also points to the possibility of multiverses (not in the sense of the many-worlds interpretations), as is clear that stories sharing common meanings can form aggregates, and that some of them might have form a very long time ago, around an initial ``seed concept.'' Just to offer another analogy, think of so-called ``shared cinematic universes'' of our recent years movie culture: each shared cinematic universe contains a growing number of films (stories) that are all meaning connected, focusing on different characters or group of characters, but all part of a same coherent and non-contradictory continuity. On the other hand, stories about characters in a given cinematic universe will never appear in another one, and if ``crossovers'' nevertheless happen (think of DC Comics' Superman possibly meeting Marvel Studios' Spider-Man), the associated stories will be usually considered to be non-canon, i.e., more abstract states of the characters involved, for instance described as alternate realities, ``what if'' scenarios, jokes and gags, dreams, etc. 

When it comes to our spatiotemporal universe and its vastness, the question of the possible presence of intelligent extraterrestrial life also arises in a natural way, also because the majority of scientists is convinced that intelligent extraterrestrials populate space, resulting in various scientific programs that over time have been funded for the search for intelligent life. As Carl Sagan used to say, in a famous science-fiction novel \cite{Sagan1985}: ``The universe is a pretty big place. If it's just us, seems like an awful waste of space.'' Space, however, would only be the tip of the iceberg of a realty whose spatiotemporal manifestation would only correspond to a thin layer of it. We can of course explore ``in width'' such layer, which certainly is a vast territory if considered from our human limited perspective, but following the view that we have expressed in the present article, there is another territory, incredibly wider, which is about exploring reality ``in depth,'' in the direction of its more abstract states. This is what physicists have begun to do when designing refined experiments about the many different quantum and relativistic effects. These experiments, and the associated efforts to describe their outcomes by means of a suitable formal language,\footnote{The unreasonable effectiveness of mathematics in the natural sciences \cite{Wigner1960} becomes all of a sudden less unreasonable if we consider that mathematics is first of all a sophisticated conceptual language and that physical entities interact in a language-mediated conceptual way.} can be seen to be our first primitive steps in learning a non-human and more universal proto-language, so perhaps it will be by exploring reality along this ``in depth'' direction that contacts with extra-terrestrial (extra-dimensional) intelligence will firstly occur, if they have not already occurred \cite{AertsSassoli2017b}. 

Let us also mention John Wheeler's famous ``it from bit'' epitome, which he used to indicate that \cite{Wheeler1989}: ``all things physical are information-theoretic in origin,'' in a participatory universe. The conceptuality interpretation completes Wheeler's account in two different ways. First of all, by extending the notion of participant, which is not limited to humans creating meaning by operating measurement devices, as the latter (and more generally, all pieces of matter) would themselves be meaning-sensitive entities able to exchange information, independently of the presence of the human consciousnesses. Secondly, by observing that ``bit,'' understood as a unit of measure in meaning exchanges, is not what combines to construct the physical entities of our spatiotemporal environment, or to more generally produce the different physical phenomena.\footnote{We cannot combine cubic meters to build a house, although its volumetric properties can certainly be expressed in such units.} What combines is not the bits of information, but the conceptual entities carrying such information, which participate in a grand conversation where the different cognitive participators, at different organizational levels, constantly exchange streams of meaningful information. So, following Wheeler's desiderata to synthesize the central point of quantum theory (and, we also add, of relativity theory) in a simple and concise statement that anyone could understand, we believe that such statement might be: \emph{the stuff the world is made of is conceptual}.

To conclude, it is important to note that the conceptuality interpretation also contains an explanation that would make our physical reality intelligible again to human pre-scientific intuition and thinking. This certainly distinguishes it from all the other interpretations, and also confers to it a highly speculative character, at least at the present state of our investigation. In that respect, it is important to mention again how crucial it is not to confuse the human conceptual realm with the conceptuality that would be inherent in our physical world. In pre-scientific times, in order to make sense of the physical entities and associated phenomena, we humans tried to psychologize them, conferring them human-like mental attributes, motivations and behaviors. According to the conceptuality interpretation, by doing so we committed a serious mistake, but at the same time we also accessed a deep insight about the physical world. The deep insight is the recognition that the latter would share with our human cultural world a same conceptual/cognitive nature; the serious mistake is about believing that physical entities and human cognitive/conceptual entities would exchange the same kind of meaning. This is the same kind of mistake we committed when we believed that planet Earth was fixed at the center of the universe, which was then reduced to a mere celestial sphere with the stars attached to it. When we escaped this ``Ptolemaic cave,'' following the Copernican revolution, we accessed an incredibly wider and richer universe. Similarly, by escaping the ``cave of our human-centered worldview,'' following the ``conceptuality revolution'' (if it will turn out to be such), we might also access an incredibly deeper and richer reality, requiring us to learn not only new languages, but also the non-human semantics attached to them.

\end{document}